\documentclass[twocolumn,amsmath,amssymb,superscriptaddress,prx,floatfix,reprint]{revtex4-2}
\usepackage{graphics,amssymb,amsmath,epsfig,color,textgreek}
\usepackage{amsthm}

\usepackage{graphicx}
\usepackage[dvipsnames]{xcolor}
\usepackage{dcolumn}% Align table columns on decimal point
\usepackage{bm}% bold math
\usepackage[colorlinks=true,citecolor=cyan]{hyperref}
\hypersetup{colorlinks=true,citecolor=cyan,linkcolor=red,urlcolor=magenta}
\usepackage{braket}
\usepackage[normalem]{ulem}
\usepackage{cancel}
\usepackage{xfrac} 
\usepackage{amsmath}
\usepackage{amssymb}
\usepackage{amsthm}
\usepackage{tikz}
\usepackage{mathtools}
\usetikzlibrary{quantikz}
\usepackage{svg}
\usepackage{lipsum}
\usepackage{booktabs}
\usepackage[capitalise,nameinlink,capitalise]{cleveref}

\crefname{section}{Sec.}{Secs.}
\crefname{table}{Tab.}{Tabs.}
\crefname{figure}{Fig.}{Figs.}
\crefname{definition}{Def.}{Defs.}
\crefname{lema}{Lem.}{Lems.}
\crefname{theorem}{Thm.}{Thms.}
\crefname{corollary}{Cor.}{Cors.}

\newtheorem*{theorem-non}{Theorem}

\newcommand{\FH}{Fermi-Hubbard\ }

\begin{document}

\author{Anjali A. Agrawal}
\affiliation{Department of Physics, North Carolina State University, Raleigh, North Carolina 27695, USA}

\author{Joshua Job}
\affiliation{Lockheed Martin, Sunnyvale, CA, 94089}

\author{Tyler L. Wilson}
\author{S. N. Saadatmand}
\author{Mark J. Hodson}
\author{Josh Y. Mutus}
\affiliation{Rigetti Computing, 775 Heinz Avenue, Berkeley, California 94710, USA}

\author{Athena Caesura}
\affiliation{Zapata AI Inc., Boston, MA 02110 USA}

\author{Peter D. Johnson}
\affiliation{Zapata AI Inc., Boston, MA 02110 USA}

\author{Justin E. Elenewski}
\affiliation{MIT Lincoln Laboratory, Lexington, Massachusetts 02421, USA}

\author{Kaitlyn J. Morrell}
\affiliation{MIT Lincoln Laboratory, Lexington, Massachusetts 02421, USA}

% Temporarily hold off on listing Kevin.
% \author{Kevin E. Obenland}
% \affiliation{MIT Lincoln Laboratory, Lexington, Massachusetts 02421, USA}

\author{Alexander F. Kemper}
\affiliation{Department of Physics, North Carolina State University, Raleigh, North Carolina 27695, USA}

\title{Quantifying fault tolerant simulation of strongly correlated systems using the Fermi-Hubbard model}

\begin{abstract}
    Understanding the physics of strongly correlated materials is one of the grand challenge problems for physics today.   A large class of scientifically interesting materials, from high-$T_c$ superconductors to spin liquids, involve medium to strong correlations, and building a holistic understanding of these materials is critical. Doing so is hindered by the competition between the kinetic energy and Coulomb repulsion, which renders both analytic and numerical methods unsatisfactory for describing interacting materials. Fault-tolerant quantum computers have been proposed as a path forward to overcome these difficulties, but this potential capability has not yet been fully assessed. Here, using the multi-orbital Fermi-Hubbard model as a representative model and a source of scalable problem specifications, we estimate the resource costs needed to use fault-tolerant quantum computers for obtaining experimentally relevant quantities such as correlation function estimation. We find that advances in quantum algorithms and hardware will be needed in order to reduce quantum resources and feasibly address utility-scale problem instances.
\end{abstract}

\maketitle

\section{Introduction}

Quantum computing has captured the interest of a
wide variety of scientists. Physicists, 
chemists,
mathematicians, 
and computer scientists
have come together to discover how this new tool
can be applied to their domain, to develop new 
algorithms that make optimal use of it, to
unveil the mathematical underpinnings of quantum
error correction, and to extend complexity classes
to incorporate quantum problems. The hardware
community is also not standing still.  The current
heavyweight contenders are platforms based on
trapped ions and non-linear superconducting resonators,
but there is rapid development of photonics-based
platforms, spin qubits, and beyond.  At this time,
however, the hardware is what has been termed
the Noisy Intermediate-Scale Quantum (NISQ) era, where
there are only a few physical qubits available,
and those that are decohere quickly and the operations
used to control them are imperfect.  The eventual
goal is to reach the Fault Tolerant era, where some
manner of error-correcting code\cite{ErrorCorrectionZoo}
is used to overcome the inevitable errors.  While
an argument can be made that we are currently in
an Early Faulty Tolerant Quantum Computing (EFTQC) eta, there remains a wide gulf to fault tolerance that
needs to be bridged by significant and sustained
hardware development.  FT quantum computers (FTQCs)
require an order of magnitude more physical qubits
than logical qubits, which comes with the associated
control hardware, potential refrigeration if low
temperatures are required, and potentially long
runtimes. 

Before we cross the gulf, it is worthwhile to ask
what the benefit is of doing so. In addition, it is
helpful to have a metric that tells us how far we
have gone.  In this work, we propose to use one of
the most studied models in condensed matter physics ---
the \FH model --- for both of these purposes.
In its prototypical form, this model represents
electrons hopping on a lattice, with an energy
penalty $U$ if two electrons are on the same
site.  The model has been studied for decades as
the simplest model for correlated systems
that shows non-trivial behavior, and in fact
exhibits many features and phases found in
nature\cite{arovas2022hubbard}. It is also extensible,
enabling a scaling in complexity as more and more
microscopic detail for a particular material are
incorporated.  As such, it makes for an ideal
generator of problems that can be used as
a metric for how far our computational effort
has grown.

In this work, we will illustrate how the \FH (FH) model can be
used to produce a scalable set of problems, ranging
from trivial to beyond the state of the art, by 
growing and extending it in several directions, and by
asking more and more of the computational tool.
At the same time, 
we will quantify some of the potential benefits of 
having an ostensible computer (quantum or not) that can 
solve this model. Naturally, these increase as the
problems become more complex, and as the solver gets
faster, and more accurate.

``Solving'' the \FH model has, in and of itself,
a utility --- an immediate, measurable benefit.
The first order effects are primarily seen in
comparing the costs (operating, energy usage,
and carbon footprint) of obtaining results with
a competing technique such as high performance 
computing. We can also quantify the benefits
of using model input to guide experimental
science, and of knowledge generation per se.
At second order, we can estimate the impact
of having a \FH model solver available in terms
of technology spillover into several sectors and human capital accumulation.

We also consider the costs of solving this
model on a quantum computer. 
Previous studies have aimed to reduce the costs of quantum primitives relevant to the Fermi-Hubbard simulation, such as Trotter step gate counts \cite{clinton2024towards}.
Other works have analyzed the costs of ground state energy estimation for the Fermi-Hubbard model \cite{yoshioka2024hunting}.
Here, we will focus on the costs of using a quantum computer to estimate properties beyond the ground state energy, which are more relevant to physical analysis. 
These quantities of
interest include static and dynamic observables
that can be directly observed with experiments.
To derive such costs we develop quantum algorithms for estimating these properties and compile circuits for these algorithms down to a logical architecture.
We then use models of the components within a modular superconducting qubit architecture to predict various costs consumed by the physical architecture to carry out the low-level processes demanded by the compiled quantum circuit.
The costs generated include basic considerations of runtime and physical qubits, but also include detailed runtime and qubit resource breakdowns of various quantum processes including T state distillation, intra-module operations, and inter-module operations.
More than providing absolute numbers for the costs of the quantum computation, these resource breakdowns point to opportunities for advancing components throughout the quantum state to drive down costs.

In what follows, we begin by describing the
\FH model and some of its properties, and how
it can be used to generate a set of problems.
We will also and list a set of
desired observables (thus indicating what ``solving''
the model means in our context).

\begin{figure*}[htpb]
    \includegraphics[clip=true,trim=0 130 0 160,width=0.95\textwidth]{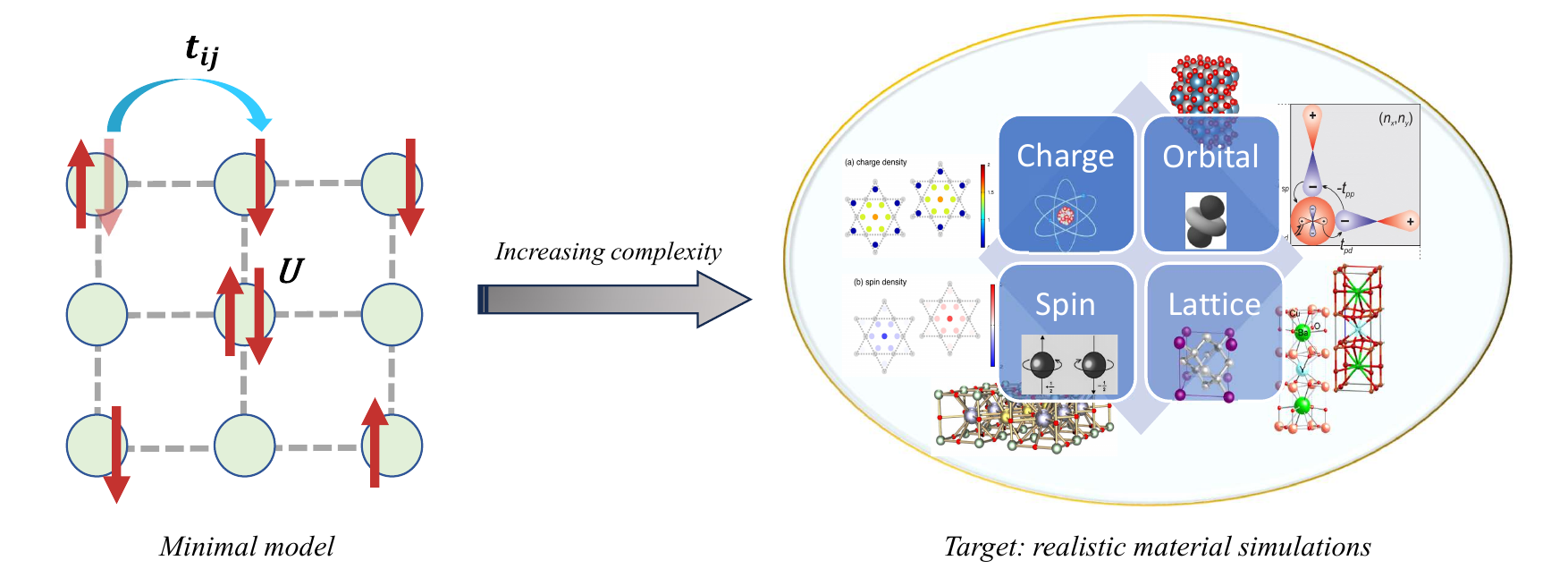}
    \caption{\textbf{The \FH model}. The base case (left) is
    a single orbital model on a two-dimensional square lattice. By increasing
    complexity through adding more degrees of
    freedom (orbital, lattice, spin, charge), a
    microscopic description of a realistic material can be achieved.}
    \label{fig:dom_FH}
\end{figure*}

\section{Prior work}
Due to the wide range of applications and interest among scholars, there have been many 
previous works in obtaining resource estimations for the Fermi-Hubbard model \cite{pathak2024requirements, Campbell_2021, Kivlichan_2020, Pathak2023quantifying}.
In Refs.~\cite{Campbell_2021, Kivlichan_2020} authors provide bounds on estimates for solving 2D Hubbard model 
with different algorithms for transmon-based superconducting hardware devices.
Logical and physical resource estimations presented in Ref.~\cite{pathak2024requirements} are calculated for one particular algorithm and 
involve Hamiltonian approximation using density matrix downfolding (DMD). 
In our work, however, we do not use any Hamiltonian approximation methods and focus on getting actual resource requirements for a given problem Hamiltonian.
In this work, we assume hardware-specific parameters and problem-dependent surface codes for error correction instead of considering general values as used in some previous calculations.

\section{The Fermi-Hubbard model}

We start with a brief overview of the
Fermi-Hubbard model as a set of problems
that can be made arbitrarily complex by
incorporating ever more degrees of freedom.
Starting from the base case --- the 2D single
orbital Fermi-Hubbard model --- one can make
a connection to materials physics by including
details about the material, turning the model
from a {\emph{drosophila}} intended to capture
just a few features to a full-fledged 
representation of the correlated model for the
material of interest.

In all cases, the \FH model as a calculation
task has a few basic parameters. Primary
is the system size $N$; as the lattice under study
grows, the calculation gets harder, but finite
size effects are ameliorated. The current state
of the art for exact diagonalization calculations is limited by the memory needed to store the Hilbert space vector; a half-filled 22-site
calculation already requires 7.5 Tb
of memory. A second parameter is the strength
of the correlations, typically denoted by a
$U$. Similar to the system size, having a
large $U$ compared to the kinetic energy can
lead to computational problems, most notably
in Quantum Monte Carlo approaches where the
sign problem rears its head in these parameter
regimes\cite{iglovikov2015geometry,hastings2016quantum,mondaini2022quantum}.  Yet, large $N$ and
large $U$ are precisely the parameter ranges
where interesting physics and potential
applications can be found.  This motivates our
choice of the \FH model as a benchmark: it has
both scalable computational complexity, as
well as potentially significant impacts on 
academics and industry.

In order to become a microscopic model for a realistic material, various degrees of freedom can be included to get closer to the real system description%\cite{chu2017van,Ikeda_2018}
; 
typically
these include charge, spin, orbital, and lattice degrees of freedom (see \cref{fig:dom_FH}).
We will start with a minimal model and build up by including these degrees to increase the system complexity aimed at getting closer to realistic material simulations.

In what follows, we will outline the base case,
the extension to a more realistic model for
materials and the extension obtained by
incorporating coupling to bosonic modes.
We will also discuss the physically-relevant
quantities one wishes to obtain from a solution
of the problem.

\subsection{Base case: the 2D, single orbital Fermi-Hubbard model}

The simplest case of the FH model,
which is minimally representative\cite{raghu2010superconductivity,arovas2022hubbard} of
a broader class of models of interest
to physicists, consist of a two-dimensional
square lattice with one orbital per site.
Its Hamiltonian is given by
\begin{align}
    \mathcal{H} &= - V_{nn}\sum_{\braket{i,j},\sigma}  (c^{\dag}_{i,\sigma} c_{j, \sigma} + h.c.)  \nonumber \\
     &+ \sum_{i} U n_{i,\downarrow} n_{i,\uparrow} - \mu \sum_{i, \sigma}  c^{\dag}_{i,\sigma} c_{i, \sigma}
     \label{eq:fh_hamiltonian}
\end{align}
where $c^\dagger_{i,\sigma}$ ($c_{i,\sigma}$)
creates (annihilates) an electron with
spin $\sigma$ at site $i$. The first
term represents the kinetic energy
in the tight-binding limit ---
$\braket{i,j}$ denotes nearest neighbour hopping with an amplitude $V_{nn}$. The interaction term induces
a Coulomb energy cost $U$ whenever two electrons are on the same site. The final
term is the chemical potential, which is
adjusted to obtain the desired density.

\subsection{From 2D Fermi-Hubbard to real materials}

As a base case, this model exhibits a
surprisingly wide variety of phases
\cite{arovas2022hubbard}, (anti-)ferromagnetism,
unconventional superconductivity, charge density
waves, and more.
On the other hand, it is a major
oversimplification of a model that is actually
representative of a material.  The Hubbard
model gained its prominence within the context
of the high-T$_\mathrm{C}$ cuprates, where
a minimal model of this type can indeed be constructed if one only considers the
$d_{x^2-y^2}$ orbital on the Cu sites,
and one ignores any multi-layers (if present).
However, the actual class of materials known
as high-T$_\mathrm{C}$ cuprates is much
larger and much more varied. Two typical materials
are YBCO (YBa$_2$Cu$_3$O$_{6+x}$) and Bi2212
(Bi$_2$Sr$_2$CaCu$_2$O$_{8+x}$), which exhibit
chain ordering in the apical oxygens in one case
and a bi-layer structure together with a larger
scale supermodulation in the other.  In other words,
while the 2D square lattice Hubbard model may
represent some common features, and thus exhibit
some commonality, it is hardly a faithful representation of any particular material.

In order to make a smooth transition to a material
model, one can extend the 2D Fermi-Hubbard model
by bringing in more of the degrees of freedom.
One enlightening example comes from the pnictide
superconductors, for example LaOFeAs. Here, the
low-energy degrees of freedom are the Fe $d_{xz/yz}$
orbitals. Indeed, one can build a model out of just these two\cite{qi2008pairing}, 
but while the Fermi surface shape is consistent,
the agreement between the model and the experiment ends there.  It is, however, relatively straightforward to extend this to a larger number of orbitals (5 per Fe site)\cite{graser2009near}, or even take into account the As degrees of freedom.  Extending the Hubbard
model in this way changes it into a multi-orbital variant,
\begin{align}
H & =  \sum_{i,j,\ell,ell'}\sum_\sigma V^{\ell,\ell'}_{i,j} c^\dagger_{i\ell\sigma} c_{j \ell' \sigma}
\nonumber \\
& +{U}\sum_{i,\ell}n_{i\ell\uparrow}n_{i\ell\downarrow}+{U}'\sum_{i,\ell'<\ell}n_{i\ell}n_{i\ell'} \nonumber\\
 &   +{J}\sum_{i,\ell'<\ell}\sum_{\sigma,\sigma'}c_{i\ell\sigma}^{\dagger}c_{i\ell'\sigma'}^{\dagger}c_{i\ell\sigma'}c_{i\ell'\sigma} \nonumber \\
 &   +{J}'\sum_{i,\ell'\neq\ell}c_{i\ell\uparrow}^{\dagger}c_{i\ell\downarrow}^{\dagger}c_{i\ell'\downarrow}c_{i\ell'\uparrow}.
 \label{eq:multiorb}
\end{align}
The terms, in order, are the kinetic energy, the intra-orbital Coulomb repulsion, the inter-orbital
Coulomb repulsion, the Hund's rule coupling, and
a pair hopping term.  

The Hamiltonian is now much
more complex as it takes into account a larger
number of degrees of freedom.  There is a concomitantly larger number of parameters as well,
which need to be determined from ab-initio
theoretical chemistry methods\cite{bauman2023coupled,solovyev2008combining,biermann2014dynamical,werner2016dynamical}, or by fitting
to experiment\cite{norman1995phenomenological}.  
Even with this more complex model, some terms remain neglected, such
as a nearest-neighbor Coulomb interaction\cite{jiang2018d,johnston2012evidence}.
The point, however, is that the model can be extended
with more and more degrees of freedom until it is
much closer to a realistic model for a particular
material, at the cost of having a much more complex problem to build and to solve.

\subsection{Including bosonic degrees of freedom}

It is well-known that bosonic modes, such as
phonon modes, spin fluctuations and more,
play an important role in describing the physics
of materials. Phonon modes play a key role
in the physics of Jahn-Teller distortions
and ferroelectricity, and show up as spectral
features in various measurements. Materials
where magnetism is present exhibit magnetic
fluctuations, and these can be measured with 
spin-resolved neutron scattering. 

In order to incorporate bosonic modes, we can
once again extend the Hamiltonian with these
degrees of freedom. Here we apply this to including
phonons in a 1-orbital model, which e.g. leads
to the appearance of kinks in the photoemission
spectra\cite{johnson2001doping,lanzara2001evidence,lee2007aspects} and has been shown to exhibit interesting interplay with superconductivity\cite{song1995electron,shen2002role,sarkar2021anomalous,ohkawa2004electron,devereaux2004anisotropic}. Note that it is
straightforward to include other bosonic degrees
of freedom in the exact same framework.
The new Hamiltonian in momentum space is\cite{Mahan,bruus}
\begin{align}
    \mathcal{H}_{\mathrm{bosons}} &= 
    \sum_{q,\lambda} \Omega_{q,\lambda} \left[ b^\dagger_{q,\lambda} b_{q,\lambda} + \frac{1}{2}\right] \\
    \mathcal{H}_{\mathrm{el-boson}} &= 
    \frac{1}{\mathcal{V}} 
    \sum_{kq\lambda}\sum_\sigma
    g_\lambda(k,q) c^\dagger_{k+q,\sigma} c_{k,\sigma} \left[
    b_{q,\lambda} + b^\dagger_{-q,\lambda}
    \right]
\end{align}
where $b^\dagger_{q,\lambda}$ ($b_{q,\lambda}$)
creates (annihilates) a phonon in
branch $\lambda$ with momentum $q$.

\subsection{Quantities of interest}

As part of the solution of the model,
one typically wishes to obtain one or more
observables (operators).  These can be
classified into static and dynamic observables.
The former involves a simple expectation value;
example of this include the local density or
magnetization, or the superconducting gap. The
latter is a non-equal time correlation function
(in the time domain), or an energy-dependent
function (in the frequency domain). Below we give
prototypical examples of each.

\subsubsection{Static observables}

Examples of static simple observables
are local, or averaged over the size of the
system, and give some indication of the
properties of the ground state. 
For this problem, typical examples are
the local density, magnetization, and staggered
magnetization (in 1D)
\begin{align}
n_i &= \sum_{\sigma} c_{i,\sigma}^{\dag} c_{i,\sigma} \\
m_i &= \sum_{\sigma} \sigma c_{i,\sigma}^{\dag} c_{i,\sigma} \\
m^s_i &= \sum_{i} (-1)^i \sigma c_{i,\sigma}^{\dag} c_{i,\sigma}.
\end{align}
To study superconducting properties, one often
relies on the $s-$ or $d-$wave projected
superconducting gap. More generally, one
projects the pair density onto a form
factor $\phi_l(k)$ corresponding to the channel of interest (here written in momentum space)
\begin{align}
    \Delta_l = \sum_k \phi_l(k) c_{k,\uparrow}^{\dag} c_{-k,\downarrow}^{\dag} c_{-k,\downarrow} c_{k,\uparrow},
\end{align}
where for $s-$wave, $\phi_s(k) =1$, for $d-$wave,
$\phi_d(k)=\cos(k_x) - \cos(k_y)$.

In addition to static averages and single-site
observables, the static correlation functions
yield information that can give hints to
underlying structure that disappears upon
averaging. Stripe structures that arise from
the interactions are one such example\cite{xu2024coexistence}.
These kinds of features can be captured by
density and magnetization correlation functions,
\begin{align}
\label{eq:dens_corr_def}
    n_i n_j &= \sum_{\sigma,\sigma'}
    c_{i,\sigma}^{\dag} c_{i,\sigma}
    c_{j,\sigma'}^{\dag} c_{j,\sigma'},
\end{align}
\begin{align}
\label{eq:mag_corr_def}
    m_i m_j &= \sum_{\sigma,\sigma'} \left(\sigma  c_{i,\sigma}^{\dag} c_{i,\sigma}\right)
    \left(\sigma' c_{j,\sigma'}^{\dag} c_{j,\sigma'}\right)
\end{align}

\subsubsection{Dynamic observables}\label{DynamicCorr}

In contrast to static observables, dynamic
observables involve the relationship between
the ground state $\ket{\psi_0}$ and the nearby
excited states. From another perspective,
dynamical observables can indicate the presence
of fluctuating order --- an order that could
lead to spontaneous symmetry breaking if
some parameter (such as temperature) is changed.
In other words, it gives hints as to whether
a particular quantity of interest is likely to
be nearby.

For a given wavefunction $\ket{\psi}$, a dynamical
observable or \emph{correlation function} is typically
of the form
\begin{align}
\label{eq:corr_function}
    \chi(t) = \langle \psi | A(t) B(0) | \psi \rangle,
\end{align}
or in Fourier domain,
\begin{align}
    \chi(\omega) = \langle\psi | A \frac{1}{\omega - \mathcal{H}} B | \psi \rangle
\end{align}
A prototypical example includes the low-energy
spectrum (where $A=B=\mathbf{1}$).
More complex correlation functions, such
as the current-current correlation function
($A=B=j$) yield direct connections to
experiments such as optical spectroscopy.
Of particular interest is the photoemission,
or particle removal spectrum
\begin{align}
    G^<_k(t) = i\braket {\psi | c^\dagger_k(t) 
    c_k(0) | \psi }.
\end{align}
This particular quantity, also known
as the lesser Green's function, can be measured
with angle-resolved photoemission spectroscopy
(ARPES)\cite{damascelli2004probing}. It measures
the single-particle spectrum, and it has
proven invaluable in the study of correlated
materials.

\section{Metrics for the solution}

\begin{figure*}
    \includegraphics[width = 0.95\textwidth]{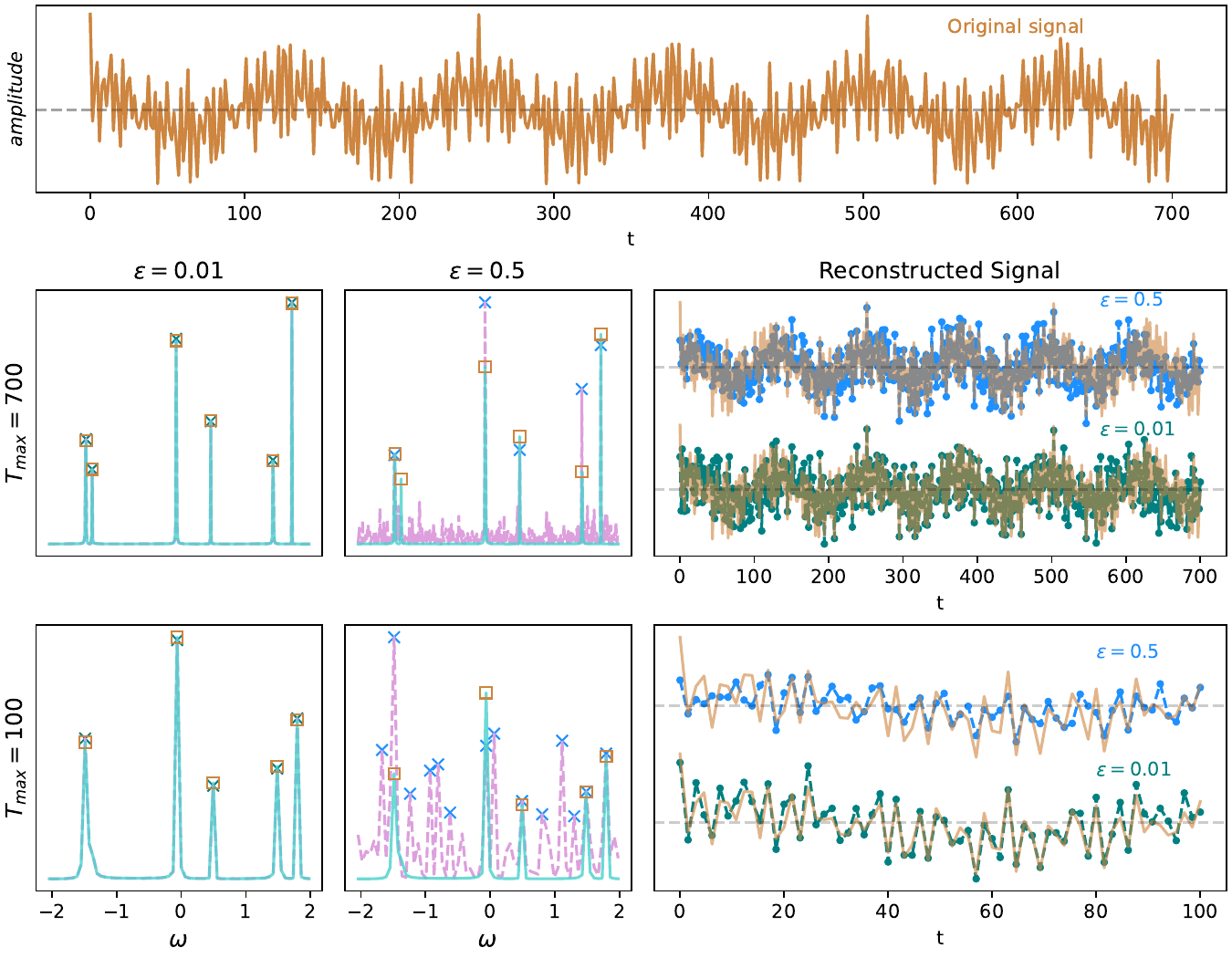}
    \caption{ Resolution test for a set of given frequencies $\omega_g = \{-1.5,-1.4,-0.05,0.5,1.5,1.8\}$ with $T_{\mathrm{max}} = \{100,700\}$ and $\epsilon = \{0.01,0.5\}$. 
    The blue lines in frequency plots correspond to Fourier transformed signal generated using specified $T_{\mathrm{max}}$ with $\epsilon=0$, and Violet dashed lines are for the signal generated with non-zero specified $\epsilon$.
    Using Python's peak finder, frequency values are identified (marked with 'x') to reconstruct the signal shown.
    Lower $T_{\mathrm{max}} = 100$ fails to resolve frequencies close to each other, (-1.5 and -1.4), and, higher $\epsilon$ makes it worse. Increasing $T_{\mathrm{max}}$ resolves the values better even with high $\epsilon$. }
    \label{fig:sampledata}
\end{figure*}

A solver taking some Hamiltonians as input and aiming at solving desired quantities of interest would require additional metrics to truncate the algorithm according to the ``quality" of the solution required.
We can describe the quality in terms of the accuracy of the solver in finding a solution with a certain precision. 
The quality of the solution can be refined further based on the quantity of interest.
This will bound what we mean by an acceptable ``solution". 

In another direction, a different metric is the amount of resources used to obtain the solution.
These resources could include the energy required for the solution and time for the solution.
Along these lines, a computation task for the solver will include specified inputs aiming to calculate a specific quantity of interest with certain accuracy and precision using some computational hours.

We can broadly classify the metrics into two categories:
\begin{enumerate}
    \item Computational resources: 
    \begin{enumerate}
        \item Time to solution
        \item Energy required
    \end{enumerate}
    \item Quality of solution: 
    \begin{enumerate}
        \item Accuracy of the solver
        \item Precision of the solution 
    \end{enumerate}
\end{enumerate}
In previous sections, we defined the inputs in terms of Hamiltonian parameters and 
desired outputs with quantities of interest for FH-type problem instances in the previous sections. 
Below, we will develop a notion of an acceptable solution for these problems using the metrics mentioned above.
While we give some analysis of static correlation function estimation in Section \ref{subsubsec:static}, our primary focus will be on the more general task of dynamical correlation function estimation.

\subsection{Accuracy for dynamical correlation functions}

\begin{figure*}[ht]
    \includegraphics[width = 0.95\textwidth]{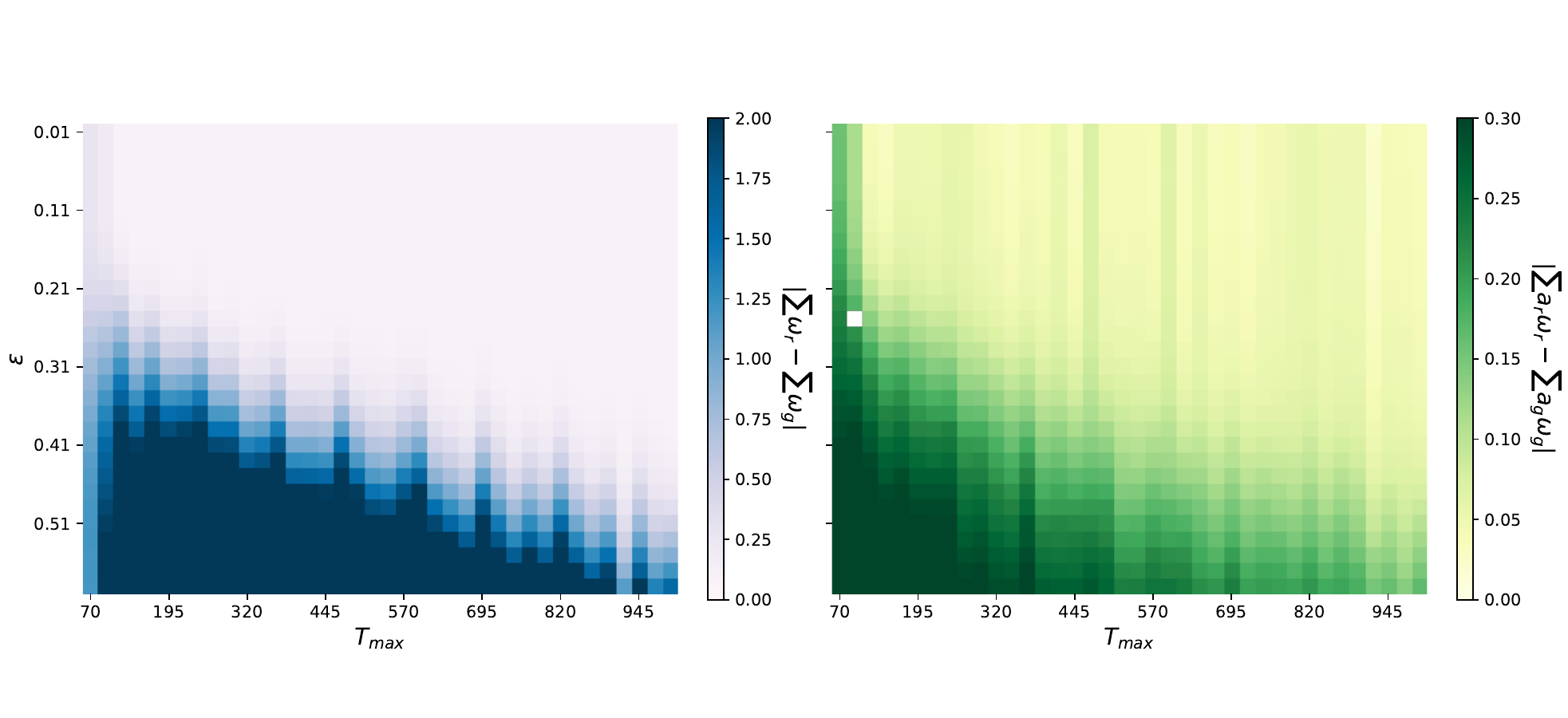}
    \caption{ Resolution test for combinations of $\epsilon$ and $T_{\mathrm{max}}$, to compare the given frequencies and amplitudes with reconstructed values. (a) Difference in the frequencies given ($\omega_g$) and reconstructed values ($\omega_r$) and (b) The difference in frequencies and
    amplitudes.}
    \label{fig:freq_amp}
\end{figure*}

For different outputs, constraints on the solution can be defined differently. 
When it comes to dynamical correlation functions,
and in particular those where the spectrum is
discrete, a reasonable measure is the accuracy
of the amplitudes and frequencies found in the
spectrum.  This is the approach we will follow
here.

As described above in \cref{DynamicCorr},
dynamic correlation functions
can be obtained by
time evolution of a state under the Hamiltonian.
In this case, the total time of evolution decides the resolution of the frequencies when plotted
via the Nyquist theorem. 
Specifically, if the energy scales defined by the Hamiltonian parameters are taken as unity, 
a total time of evolution  $T_{\mathrm{max}} = 2 \pi\times x$ where $x \in \{1,10,100,1000,...\}$
yields a corresponding resolution of $\{1,0.1,0.01,0.001,...\}$. An additional parameter
that can vary is the sampling rate $dt$,
which bounds the maximum observable frequency
as $\omega_{\mathrm{max}} = \pi/dt$.
Finally, for data coming from a quantum computer,
we can expect a given accuracy. That is,
if the accuracy of the solver is $\epsilon$, 
for an accurate data value $a$
the data points obtained lie anywhere in $(a-\epsilon, a+\epsilon)$.

Combining these, it is straightforward to see that having a larger $T_{\mathrm{max}}$ and smaller $\epsilon$ produces a more accurate set
of amplitudes and frequencies.  This is further
compounded by the notion of Fourier filtering
--- if sufficient data is available, frequencies
can reliably be obtained even in the presence
of noise\cite{moitra2015super}.

In Fig.~\ref{fig:sampledata},
we will use an example to show how these factors interplay to set a resolution. 
We use 6 fixed frequencies, two values of $T_{\mathrm{max}}$, and two noise values 
$\epsilon$. To the original signal, we add
the worst case noise: the signal at each time point gets replaced $a \rightarrow a \pm \epsilon$, with the sign randomly chosen.
We perform a straightforward analysis to
obtain a measure of quality. We Fourier
transform the data, and extract the peak 
amplitudes and frequencies. 
In Fig.~\ref{fig:sampledata} we also compare the 
signal that is reconstructed from the peaks
in the noisy Fourier transform to the similarly reconstructed clean signal. As is clear
from both the power spectra as well as the
reconstructued signals, the best results are
obtained with large $T_{\mathrm{max}}$ and
small $\epsilon$ --- loosening either constraint
leads to errors in amplitudes, frequencies,
or both.  

It is, of course, clear that reconstructing
the signal purely from its peaks is insufficient
for a full reconstruction of the input data.
The reason for this approach is that it matches
what one might do when faced with noisy
experimental data.  Typically, a noise floor
of some type is present, below which there
is no coherent signal. Whether this noise
floor is due to physics or experimental details,
it is not straightforward to incorporate this
in the analysis, and is thus typically neglected.
We have attempted to capture this modus
operandi here.

To make a more quantitative statement,
we have repeated the above procedure with
randomly sampled frequencies and amplitudes.
We produced a signal with 6 frequencies randomly chosen from $\omega \in \{-2, -1.9, \dots 2.0\}$, 
and corresponding amplitudes are chosen from $a \in \{0.4, 0.45, \dots 1.0\}$.
We vary $T_{\mathrm{max}}\in \{ 70, 95, \dots 1000\}$, but keep
$dt$ fixed at $1.5$ (resultings in $\omega_{\mathrm{max}}\ \sim 2$,
and vary $\epsilon \in \{0.01, 0.03, \dots 0.6\}$.
Similarly to the above, we add random noise
of magnitude $\epsilon$ by replacing
the true value $a \rightarrow a \pm \epsilon$.

Based on the sampling of $\epsilon$, a different ``noisy" signal is produced each time. 
Fourier transformation of these signals produces peaks corresponding to the frequencies with some amplitudes.

As a measure, we use two quantities to compare.
The first compares just the frequencies
found in the clean signal and the
noisy signal;
the second also takes into account the amplitudes.
We considered $30$ different frequencies and 
corresponding amplitudes all chosen from the set 
mentioned above;
since the sampled signal with $\epsilon$ generates a different signal each time, we  produced $100$ curves for each value of 
$\epsilon$, $T_{\mathrm{max}}$, and, set of frequencies and amplitudes.
The results are shown in \cref{fig:freq_amp}.
We observe an expected dependence showing
better results at larger $T_{\mathrm{max}}$ and higher accuracy at smaller $\epsilon$.

The above analysis is inspired by the Nyquist sampling theory, where
the sampling rate is chosen such that the minimum requirements are met.
There exist many frequency estimation algorithms and signal post-processing techniques that aim to extract relevant information from a given input signal, and
some of these techniques can also handle noisy data and post-process it to produce a cleaner version, 
potentially reducing the required total time $T_{max}$. 
For instance, the MUSIC algorithm \cite{stocia1989MUSIC} identifies the noise subspace and uses it to filter the signal. 
They are nevertheless susceptible to noise, and in particular have difficulties telling apart frequencies
when the noise increases above a certain point\cite{moitra2015super}. 
Incorporating MUSIC and/or other techniques is straightforward, and simply affects the multiplicative factors
in the analysis that follows by adjusting $T_{max}$ and the sampling rate.

\section{Quantum Resource estimation}

\subsection{Quantum Algorithms}

\subsubsection{Static Correlation Functions: Ground State Property Estimation}
\label{subsubsec:static}

Here we describe the quantum algorithms that can be used to estimate the density-density correlation function  
\begin{align}
\braket{\psi_{\textup{gs}}|n_in_j|\psi_{\textup{gs}}}
\end{align}
and the magnetization correlation function
\begin{align}
\braket{\psi_{\textup{gs}}|m_im_j|\psi_{\textup{gs}}},
\end{align}
where $\bra{\psi_{\textup{gs}}}$ is the ground state.
Although we do not generate detailed resource estimates for this task, we have chosen to detail the algorithm for solving it and providing analytic expressions for its cost that could be used in future work.
The first step in developing the quantum algorithm is to encode the target quantities as functions of squared amplitudes.
The enables the quantities to be estimated using a quantum amplitude estimation algorithm.

We exploit the fact, shown in Eqs. \ref{eq:dens_corr_def} and \ref{eq:mag_corr_def}, that both the density-density and magnetization observables can be expressed as the sum of $W_{ij\sigma\sigma'}=c_{i,\sigma}^{\dag} c_{i,\sigma}c_{j,\sigma'}^{\dag} c_{j,\sigma'}$, which are projectors.
By estimating $\braket{\psi_{\textup{gs}}|W_{ij\sigma\sigma'}|\psi_{\textup{gs}}}$ for all four combinations of $\sigma, \sigma'$, we can linearly combine the estimates to obtain an estimate of either quantity of interest.
The estimation of such quantities can be achieved as follows.

First, we use the fact that $W_{ij\sigma\sigma'}$ is an orthogonal projector to write $\ket{\psi_{\textup{gs}}}$ as a superposition of two states, $\ket{\phi_+}\propto W_{ij\sigma\sigma'}\ket{\psi_{\textup{gs}}}$ and $\ket{\phi_-}\propto (\mathbb{I}-W_{ij\sigma\sigma'})\ket{\psi_{\textup{gs}}}$, that are orthogonal to one another,
\begin{align}
\ket{\psi_{\textup{gs}}}&=W_{ij\sigma\sigma'}\ket{\psi_{\textup{gs}}}+(\mathbb{I}-W_{ij\sigma\sigma'})\ket{\psi_{\textup{gs}}}\\
&=\cos{\theta}\ket{\phi_+}+\sin{\theta}\ket{\phi_-},
\end{align}
where the amplitude of interest is $\cos^2{\theta}=\braket{\psi_{\textup{gs}}|W_{ij\sigma\sigma'}|\psi_{\textup{gs}}}$.
Then, to arrive at a setting for amplitude estimation, we first define $U_{gs}$ as the unitary that prepares the ground state, $R_0$ as the unitary that applies a reflection about $\ket{0^n}$, and $R_{ij\sigma\sigma'}=2W_{ij\sigma\sigma'}-\mathbb{I}$.
Then, we note that multiple applications of the Grover iterate $G=U_{gs}R_0U_{gs}^{\dagger}R_{ij\sigma\sigma'}$ to $\ket{\psi_{\textup{gs}}}$ will yield
\begin{align}
G^kU_{gs}\ket{0^n}=\cos{((2k+1)\theta)}\ket{\phi_+}+\sin{((2k+1)\theta)}\ket{\phi_-}.
\end{align}
Measurement of $n_in_j$ will return outcome $1$ with probability
\begin{align}
    \textup{Pr}(1|k)=\cos^2((2k+1)\theta).
\end{align}
Amplitude estimation algorithms such as \cite{grinko2021iterative} use the ability to sample from this distribution for varying $k$ to estimate $\cos\theta$. 
The iterative amplitude estimation algorithm \cite{grinko2021iterative} ensures that the estimate, $\hat{a}$, satisfies $|\hat{a}-\braket{\psi_{\textup{gs}}|n_in_j|\psi_{\textup{gs}}}|\leq \epsilon$ with probability greater than $1-\delta_S$ by using no more than
$\pi/8\epsilon$ calls to the Grover iterate per circuit and, including the circuit repetitions, no more than a total of
\begin{align}
\label{eq:amp_est_iterates}
\frac{50}{\epsilon} \log \left( \frac{2}{\delta_S} \log_2 \left( \frac{\pi}{4\epsilon} \right) \right)
\end{align}
calls to the Grover iterate.
While this result is analytically derived, we note that recent work \cite{labib2024quantum} established empirical performance metrics for the task of amplitude estimation that suggest more than an order-of-magnitude reduction in the number of calls to the Grover iterate.

What remains is to describe how the problem-specific operations are implemented and to give their associated costs.
The simplest operation, of those described is the reflection $R_{ij\sigma\sigma'}$.
Using the Jordan-Wigner encoding, the  $R_{ij\sigma\sigma'}$ operation can be implemented using just a few Clifford gates.
The more substantial operation is $U_{\textup{gs}}$, a coherent preparation of the ground state.
From \cite{lin2020near}, we have that, given a $(\alpha, m, 0)$-
block-encoding\footnote{As explained in \cite{gilyen2019quantum}, the three values $(a, n, \delta)$ that characterize a block encoding unitary $U_B$ of a matrix $B$ correspond to the subnormalization, ancilla qubit, and error parameters in the expression $|B-\alpha\bra{0^n}U_B\ket{0^n}|\leq \delta$.} of $H$ and a unitary $U_I$ that prepares an initial state with ground state overlap $\gamma\leq|\braket{\psi_\textup{gs}|U_I|0^n}|$, we can construct a $1-\delta_{\textup{gs}}$-accurate ground state preparation unitary $U_{\textup{gs}}$ using
\begin{align}
O\left(\frac{\alpha}{\gamma \Delta} \log\left(\frac{1}{\gamma \delta_{\textup{gs}}}\right)\right)    
\label{eq:gsp_bigo}
\end{align} 
calls to $U_H$, where $\Delta$ is a lower bound on the spectral gap of $H$.
Defining $C$ to be the constant factor of Eq. \ref{eq:gsp_bigo} we have that the maximum number of calls to $U_H$ per circuit is
\begin{align}
    \frac{\pi C\alpha}{4\gamma \Delta\epsilon} \log\left(\frac{1}{\gamma \delta_{\textup{gs}}}\right)
\end{align}
Given an overall failure tolerance of $\bar{\delta}$, we will allocate this failure tolerance evenly across statistical failure $\delta_S=\bar{\delta}/2$ and circuit failure $\delta_C=\bar{\delta}/2$.
We will then allocate the circuit failure tolerance evenly among ground state preparation failure rate $\delta_{\textup{gs}}=\bar{\delta}/8$, synthesis failure rate $\delta_{\textup{syn}}=\bar{\delta}/8$, distillation failure rate $\delta_{\textup{dist}}=\bar{\delta}/8$, and data failure rate $\delta_{\textup{data}}=\bar{\delta}/8$.
Defining $T_{H}(\delta_{\textup{syn}})$ (where we have made explicit that this count depends on the synthesis failure tolerance) and $T_0$ to be the number of T gates needed to implement $U_H$ and $R_0$, respectively, the maximum total number of T gates per circuit is
\begin{align}
    \frac{\pi C\alpha}{8\gamma \Delta\epsilon} \log\left(\frac{1}{\gamma \delta_{\textup{gs}}}\right)(2T_H+T_0)
\end{align}
and the total number of T gates that need to be run throughout the amplitude estimation algorithm is
\begin{align}
    \frac{50C\alpha}{\gamma \Delta \epsilon} \log \left( \frac{4}{\bar{\delta}} \log_2 \left( \frac{\pi}{4\epsilon} \right) \right) \log\left(\frac{8}{\gamma \bar{\delta}}\right)(2T_H(\bar{\delta}/8)+T_0).
\end{align}
The full task (either density-density or magnetic correlation function) requires estimating four of these amplitudes for each of the $N(N-1)/2$ pairs of sites, giving $2N(N-1)$ total amplitude estimation tasks.
It is possible to improve this scaling.
The techniques introduced in \cite{huggins2022nearly} enables estimating multiple expectation values within the same quantum circuit.
This parallelization would improve the scaling to $O(\sqrt{2N(N-1)})$, which might be beneficial in the large-$N$ setting, although we do not explore this here.

While we have provided the cost analysis of this task here, we will not make explicit resource estimates using this. 
Instead we will focus on a different task of estimating dynamic correlation functions.
This task and the algorithm used is described in the following subsection.
Quantum resource estimates for this task are then presented in Section \ref{subsec:qre}.

\subsubsection{Dynamic Correlation Function Estimation}
Here we describe the quantum algorithm used to estimate the dynamic correlation function calculation.
As given in Eq. \ref{eq:corr_function}, the relevant computational task is to provide an estimate of the quantity $
\chi(t)=\bra{\psi}A(t)B\ket{\psi}$,
where $A(t)=e^{iHt}Ae^{-iHt}$,
over a range of different times $t$ to within a specified estimation error $\epsilon$ and failure rate tolerance $\delta$.
The operators $A$ and $B$ are assumed to be decomposable into a linear combination of a few Pauli operators $P_{\nu}$,
\begin{align}
    A = \sum_{\nu} q_{\nu}P_{\nu}\nonumber \\
    B = \sum_{\nu} q'_{\nu}P_{\nu}.
\end{align}
In particular, for the case of creation and annihilation operators (typically of interest), only two Pauli operators are needed, enabling $g(t)$ to be expressed as a linear combination of just four expectation values involving Pauli operators.
We will focus on the cost of estimating one of these terms and consider the quantity of interest to be
\begin{align}
g(t)=\bra{\psi}e^{iHt}P_{\nu}e^{-iHt}P_{\nu'}\ket{\psi}.
\end{align}
The estimation error is defined as the absolute value of the difference between the estimate and the true value of $g(t)$, while the failure rate tolerance is the probability with which the error is allowed to exceed $\epsilon$.

\begin{figure}[htpb]
\scalebox{0.8}{
\begin{quantikz}[transparent]
\qw & \gate{H} & \ctrl{1} & \qw & \ctrl{1} & \qw & \qw & \qw \\
\qw & \gate[nwires=2,wires=3]{U_{SP}} & \gate[nwires=2, wires=3]{P_{\nu'}} & \gate[nwires=2,wires=3]{U(t)} & \gate[nwires=2, wires=3]{P_{\nu}} & \gate[nwires=2,wires=3]{U(-t)} & \gate[nwires=2,wires=3]{U_{SP}^{\dagger}} & \qw \\
\vdots &&&&&&& \vdots \\
\qw & \qw & \qw & \qw & \qw & \qw & \qw & \qw \\
\end{quantikz}
}
    \caption{The core quantum circuit used in the dynamic correlation function estimation. This circuit prepares a state for which the overlap with $\ket{0}$ is a known function of $g(t)$. Depending on whether the real or imaginary part of $g(t)$ is being estimated, either a Hadamard gate $H$ or a Hadamard and phase gate $HS$ are applied.}
    \label{fig:qae_state_prep}
\end{figure}
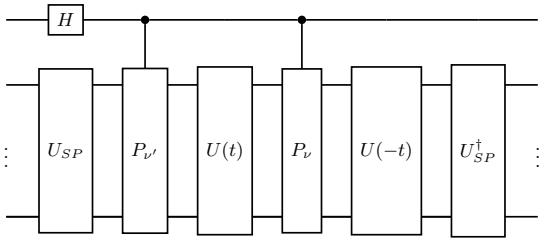

An approach to using a quantum computer to estimate $g(t)$ is to encode the real and imaginary parts of $g(t)$ into quantum amplitudes and then use a quantum algorithm for the amplitude estimation task to estimate these separately, 
similar to the case of static correlation function estimation.
We can encode each part of $g(t)$ into a quantum amplitude as follows.
First, assume that we are able to implement unitary operations $U_{SP}$, $U(t)$, $c$-$P_{\nu}$, satisfying $c$-$P_{\nu'}$, satisfying 
\begin{align}
U_{SP}\ket{0}^n&=\ket{\psi}\\
U(t)&=e^{-iHt}.
\end{align}
Then, we have that
\begin{align}
\frac{1+\operatorname{Re}(g(t))}{2}
&= |\bra{+}V(t) \ket{0}\ket{0^{n}}|^2\nonumber\\
\frac{1-\operatorname{Im}(g(t))}{2}
&= |\bra{i}V(t) \ket{0}\ket{0^{n}}|^2,
\label{eq:qamp}
\end{align}
where the structure of $V(t)=U_{SP}^{\dagger}U(-t)c$-$P_{\nu} U(t)c$-$P_{\nu'}(H\otimes U_{SP})$ is defined in Figure \ref{fig:qae_state_prep} and $\ket{i}=(\ket{0}+i\ket{1})/\sqrt{2})$.
% The imaginary part of $g(t)$ is obtained by using $\ket{i}$, prepared by $HS$, instead of $\ket{+}$.
The quantity on the left-hand-side of Eq. \ref{eq:qamp} is the squared amplitude of the state $V(t)\ket{0}\ket{0^{n}}$ on the subspace marked by $-H\otimes \mathbb{I}_{n}$ for the real case and $-SHS^{\dagger}\otimes \mathbb{I}_{n}$ for the imaginary case, where $S = \begin{pmatrix}
1 & 0 \\
0 & i
\end{pmatrix}$ is an $S$ gate.
Therefore, we can estimate the left-hand side (and thus the real and imaginary components of $g(t)$) using quantum amplitude estimation.
The circuit used in quantum amplitude estimation is based on the Grover iterate $W=(-H\otimes \mathbb{I}_{n})(V(t)R_0V(t)^{\dagger})$ for the real part (and similarly for the imaginary part), where $R_0$ is the reflection about the all-zero state.

To estimate either the quantity $\operatorname{Re}(g(t))$ or $\operatorname{Im}(g(t))$ to within estimation error $\epsilon$ below failure rate $\delta$, we can use the iterative amplitude estimation algorithm of \cite{grinko2021iterative} as described in the previous subsection with costs given in Eq. \ref{eq:amp_est_iterates}.
These costs refer to the number of circuit repetitions and the number of Grover iterates per circuit.
A key feature of this costing used in the following section is that these costs only depend on the desired estimation error and failure tolerance. Accordingly, the costs can be established independently of time $t$ and Hamiltonian $H$.

\subsection{Resource Estimates For Dynamic Correlation Function}
\label{subsec:qre}

In this section we present quantum resource estimates for the task of estimating $g(t)$ for a single time $t$.
As described in the previous section, the broader task of estimating an approximation of the dynamical correlation function $\chi(t)$ (see Eq. \ref{eq:corr_function}) can be solved by estimating $g(t)$ over different times and different Pauli operators.
These different estimations can be implemented sequentially or in parallel, incurring either a time or space multiplicative resource overhead.
Accordingly, we will focus on estimating the quantum resources required to estimate $g(t)$ for a single time.

We assume that physical resource cost is dominated by the cost of implementing the time evolution operator $U(t)$ relative to the operations $U_{SP}$, $c$-$P_{\nu}$, etc. (see \cref{fig:qae_state_prep}) \footnote{The costs of the $U_{SP}$ and $U(t)$ are far more expensive than the controlled-Pauli operators. Furthermore, there exist numerous heuristic methods for state preparation $U_{SP}$ \cite{stanisic2022observing, sugisaki2022adiabatic, fomichev2023initial} that aim to approximate the ground state with few quantum gates and future improvements are expected. Without concrete estimates of the costs of these heuristic methods, we will assume that the number of gates used to implement $U_{SP}$ is far less than that for $U(t)$. However, we leave it to future investigations to explore this assumption.} we present detailed resource estimates of the total cost of this operation. This costing takes into account the number of invocations of $U(t)$ for amplitude estimation as well as the estimated number of shots required for accurate expectation value estimates.

\subsubsection{Full Dynamic Correlation Circuit Parameters}

To arrive at costs for running quantum computations to estimate the dynamic correlation function, we must specify the requirements.
We choose the allowable error  according to the results of Figure \ref{fig:freq_amp}
as $\epsilon = 0.01$ as warrants sufficient resolution of the peaks.
We choose a failure tolerance of $\delta=0.001$, noting that the logarithmic dependence in Eq. \ref{eq:amp_est_iterates} ensures that the resource costs will not vary significantly as this is changed over several orders of magnitude.
Then, using 
$\pi/8\epsilon$ Grover iterates per circuit and a total number of Grover iterates given in Eq. \ref{eq:amp_est_iterates},
the quantum amplitude estimation costs are:
\begin{itemize}
\item Number of shots: 671
\item Applications of $U(t)$ per circuit: 156
\item Failure tolerance per circuit: $1.5\times10^{-6}$
\end{itemize}
Letting $p_{\text{qsp}}$ be the success probability of one invocation of the $U(t)$ sub-circuit, the per-circuit failure tolerance can be met by ensuring
$$ (1-1.5\times 10^{-6}) \leq p_{\text{qsp}}^{156}.$$
The smallest value of $p_{\text{qsp}}$ that satisfies this constraint is,
$p_{\text{qsp}}^{156} = 0.9999999904$ which is the value used for the resource estimates.

\subsubsection{Time Evolution Circuits}
Circuits to carry out time evolution were based on Quantum Signal Processing \cite{low2017optimal,low2019hamiltonian} and prepared by \verb|pyLIQTR| version 0.3.1 \cite{pyliqtr}. Define the precision to which the $U(t)$ approximates the time evolution $e^{-iHt}$ by $\epsilon_{\text{qsp}}$. From \cite{low2017optimal} the success probability of a QSP circuit is at least $1-2 \epsilon_{\text{qsp}}$, so for our circuits we set
$$\epsilon_{\text{qsp}} = \frac{1-p_{\text{qsp}}}{2} = 4.81 \times 10^{-9}.$$

Circuits were decomposed into ``widgets'' using Rigetti's Resource Estimation (RRE) widgetization tool \cite{RRE}. This tool recursively decomposes the nested operations in the circuits generated by \verb|pyLIQTR|, and constructs a sub-circuit dependency graph. During decomposition, repeated sub-circuits are identified and decomposition for these sub-circuits is skipped and an indexing of unique sub-circuits is constructed. Once the sub-circuit dependency graph is constructed, one may generate the full circuit from a time ordered sequence of the unique sub-circuit index. Moreover, for downstream processing, one may recombine these sub-circuits into fixed width sub-circuits, "widgets", such that they all have near uniform depth and execute the full circuit when played sequentially. These widgets are then passed into RRE's resource estimation tool.

\subsubsection{Physical Architecture and Error Correction}
Estimates were carried out assuming a two fridge measurement based quantum computer (MBQC), with each fridge containing one million physical superconducting qubits. The physical qubits in a fridge are allocated to one of three functions; computational logic qubits, ancilla qubits for connecting computational qubits, or T state distillation factories. All qubits that are not involved in T state distillation are denoted ``Bus" in Figures.

In this MBQC, graph-state compilation framework, the two fridge design allows one fridge to prepare the graph state for the next widget, while the other fridge performs measurements which consume the graph state for the current widget. 

Physical qubits were assumed to have an error rate of 0.001, and error correction was carried out using a rotated surface code. The hardware assumptions and error correction configuration was identical to those in \cite{RRE}. For additional details on system configuration, the graph-state compilation framework, and RRE's estimation tool outputs, see \cite{RRE}.

\subsubsection{Results}
The RRE estimation tool is able to provide over 60 resource estimates and circuit diagnostic quantities, including the total time evolution required to calculate dynamic correlation functions. These results were for Fermi-Hubbard Hamiltonians of the form \cref{eq:fh_hamiltonian} where $U=2, V_{nn}=1, \mu=1$, with the lattices sizes ranging from $2 \times 2$ to $7 \times 7$. 

\begin{figure}[!htpb]
    \includegraphics[width=\columnwidth]{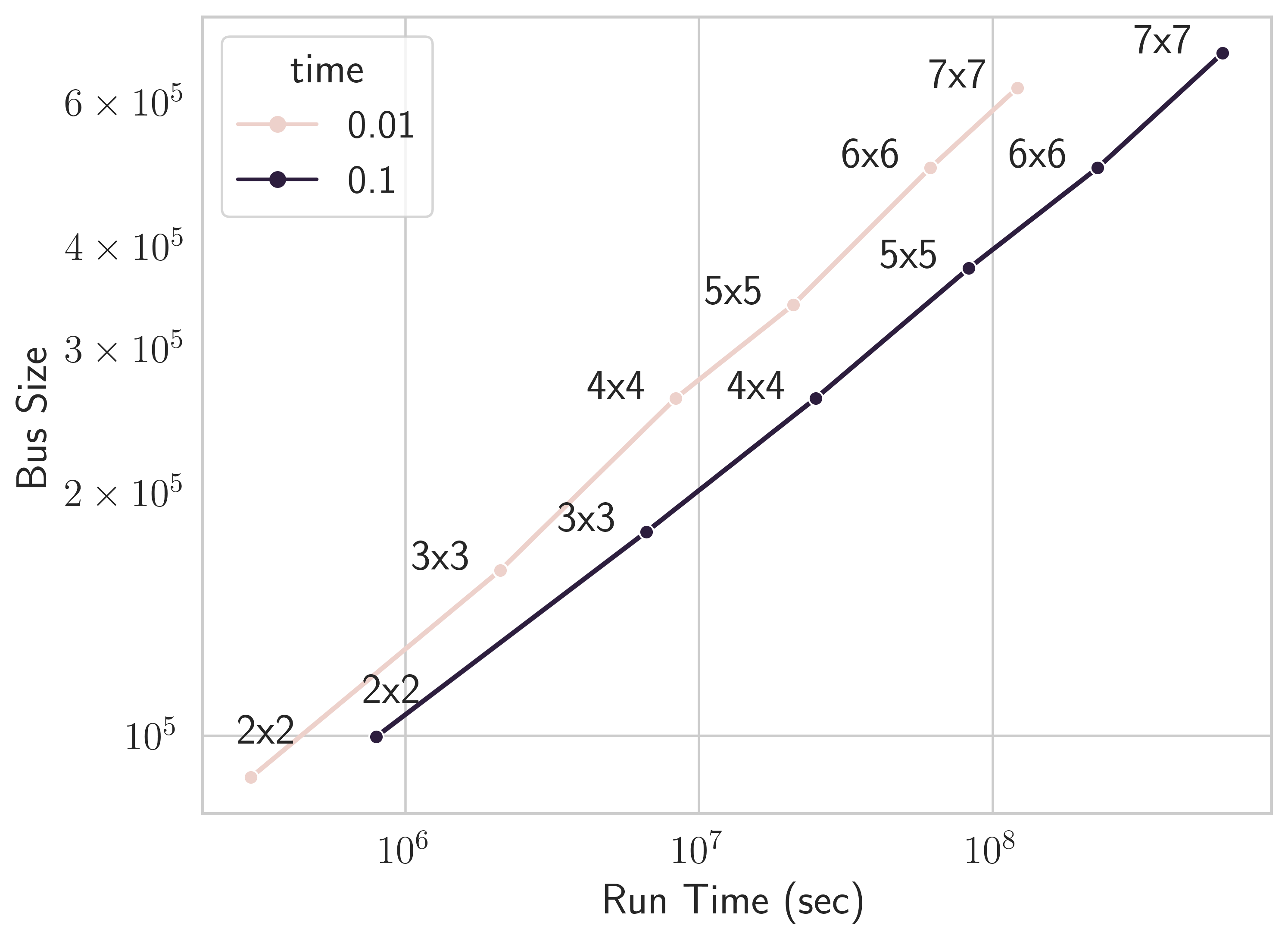}
    \caption{Physical resource scaling relationship for the time evolution portion of calculating dynamic correlation functions. The Hamiltonian considered is the Fermi-Hubbard model \cref{eq:fh_hamiltonian} with $U=2, V_{nn}=1, \mu=1$ and various lattice sizes. The bus size represents all physical qubits that are not involved in T state distillation. Run time denotes total run time of the quantum computer given in seconds. This run time accounts for all circuit invocations and shots.}
    \label{fig:bus_vs_runtime}
\end{figure}

In Fig.~\ref{fig:bus_vs_runtime} we show the scaling relationship between the number of bus qubits and total quantum run time.

\begin{figure}[!htpb]
    \includegraphics[width=\columnwidth]{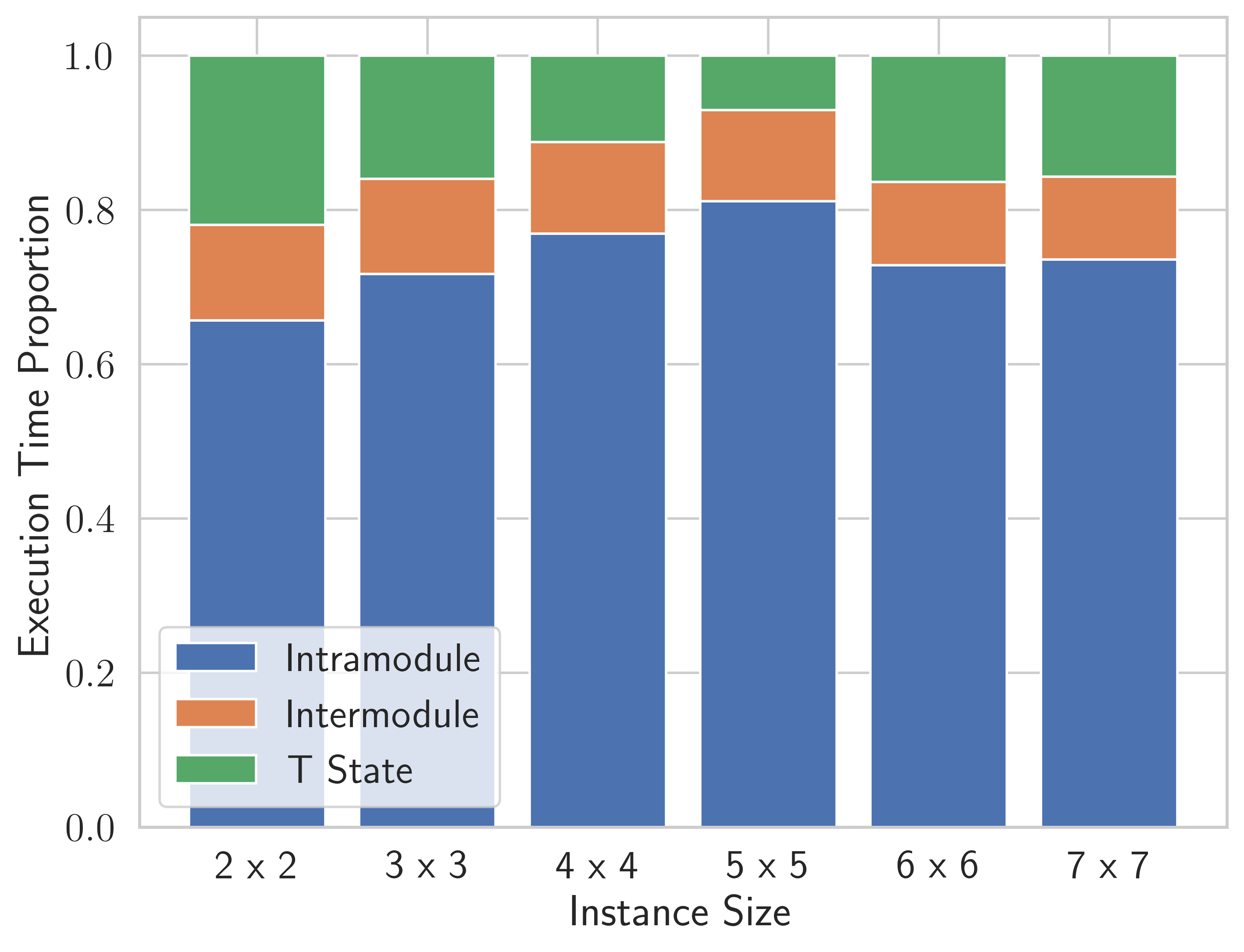}
    \caption{Proportion of quantum computer compute time allocated to T state distillation and injection, intermodule communications, and intramodule operations for time evolutions. States are evolved up to time $t=0.01$.}       \label{fig:time_alloc_t.01}
\end{figure}

\begin{figure}[!htpb]
    \includegraphics[width=\columnwidth]{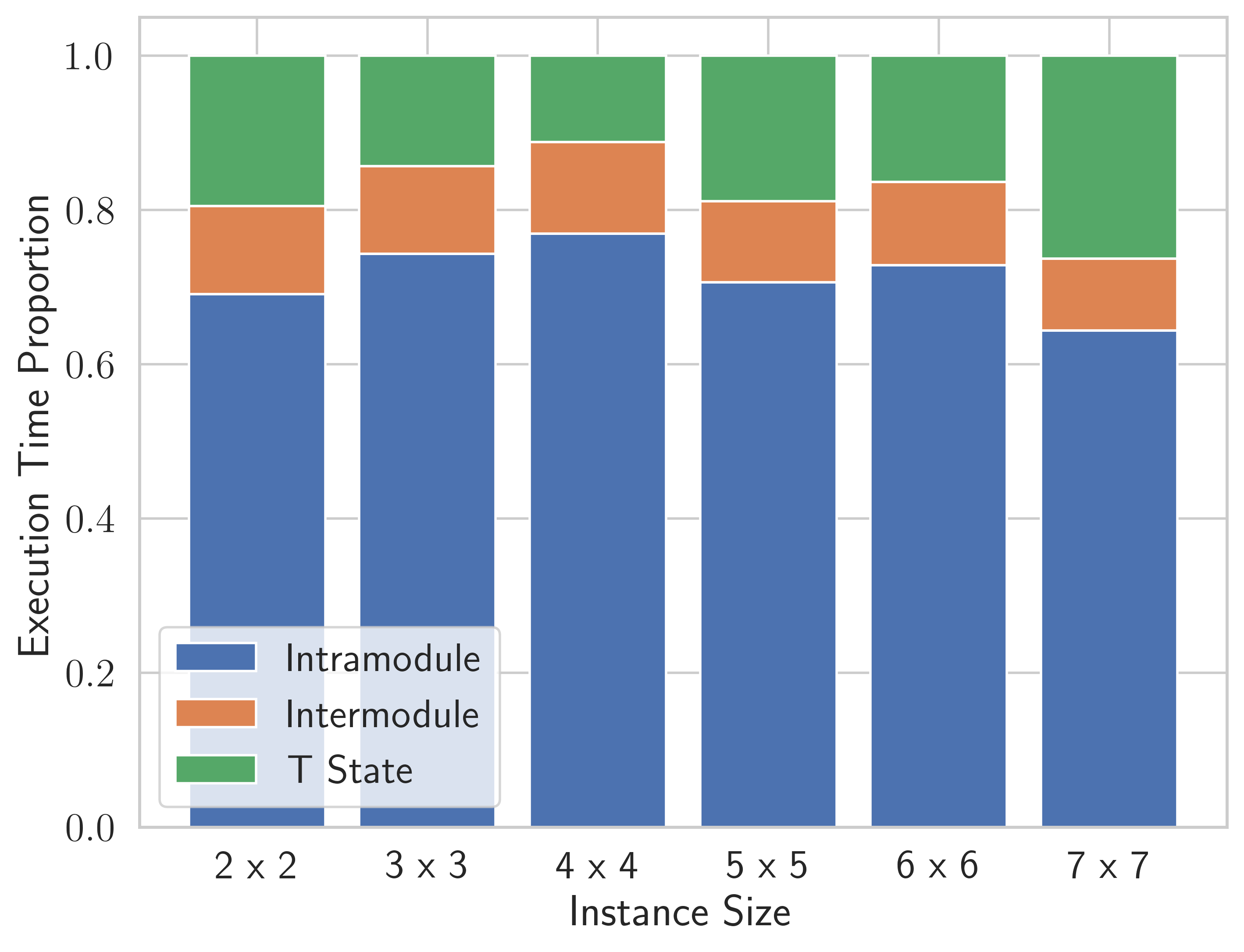}
    \caption{Proportion of quantum computer compute time allocated to T state distillation, intermodule communications, and intramodule operations for time evolutions. States are evolved up to time $t=0.1$.}       \label{fig:time_alloc_t.1}
\end{figure}

In Figs.~\ref{fig:time_alloc_t.01} and \ref{fig:time_alloc_t.1} we show the proportion of quantum computer compute time allocated to T state distillation and injection, intermodule communications, and intramodule operations. Intramodule operations are all operations which happen inside of an individual fridge such as graph state creation, consumption, teleportation, etc. 

\begin{figure}[!htpb]
         \includegraphics[width=\columnwidth]{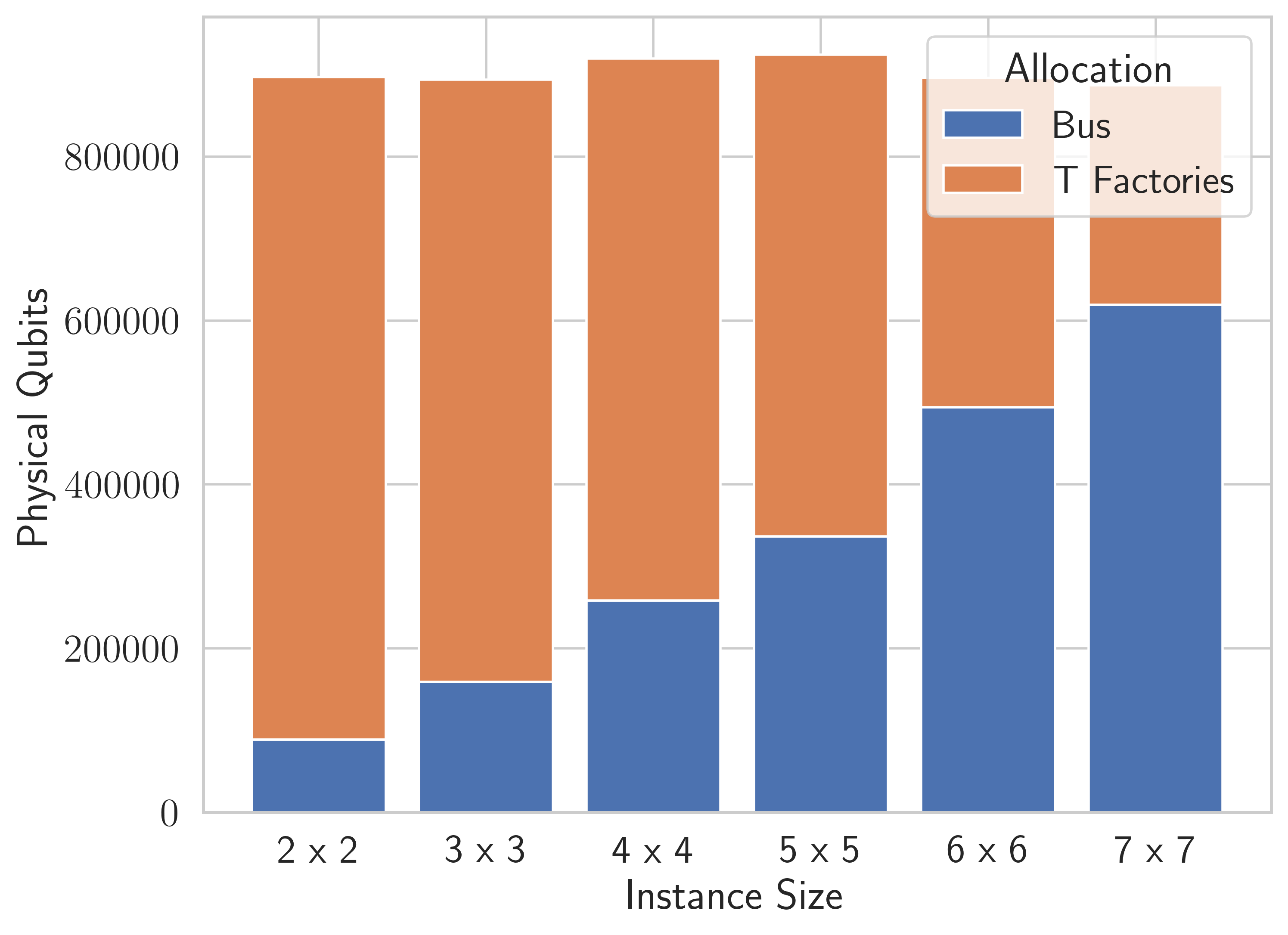}
         \caption{Physical qubits allocations for T state factories and everything else (bus) during time evolution. States are evolved up to time 0.01}
         \label{fig:qubit_alloc_t.01}
\end{figure}

\begin{figure}[!htpb]
         \includegraphics[width=\columnwidth]{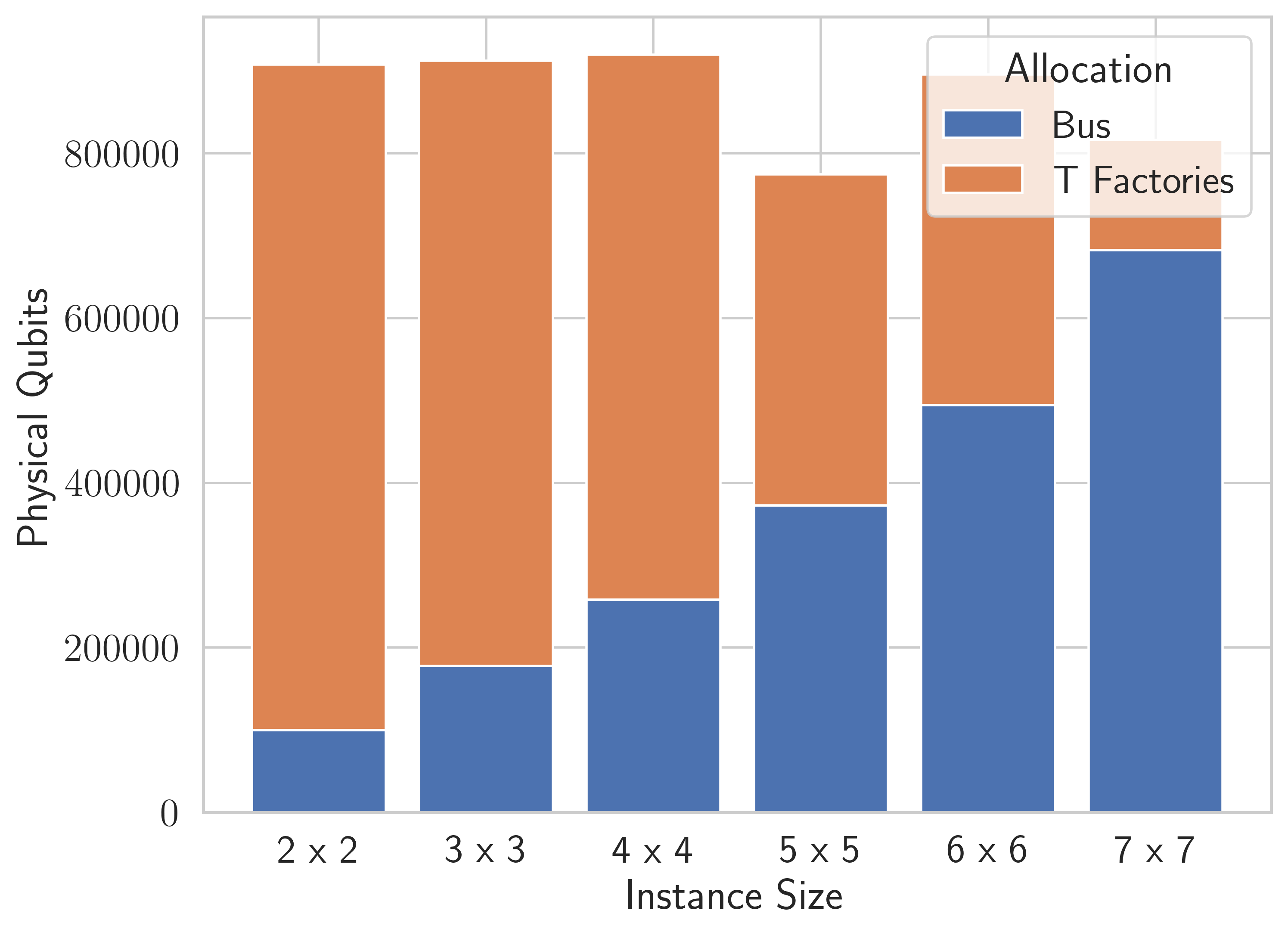}
         \caption{Physical qubits allocations for T state factories and everything else (bus) during time evolution. States are evolved up to time 0.1}
         \label{fig:qubit_alloc_t.1}
\end{figure}

Similarly, in Figs.~\ref{fig:qubit_alloc_t.01} and \ref{fig:qubit_alloc_t.1} we show the physical qubits that are T state factories or all other operations (bus).

\begin{figure}[!htpb]
    \includegraphics[width=\columnwidth]{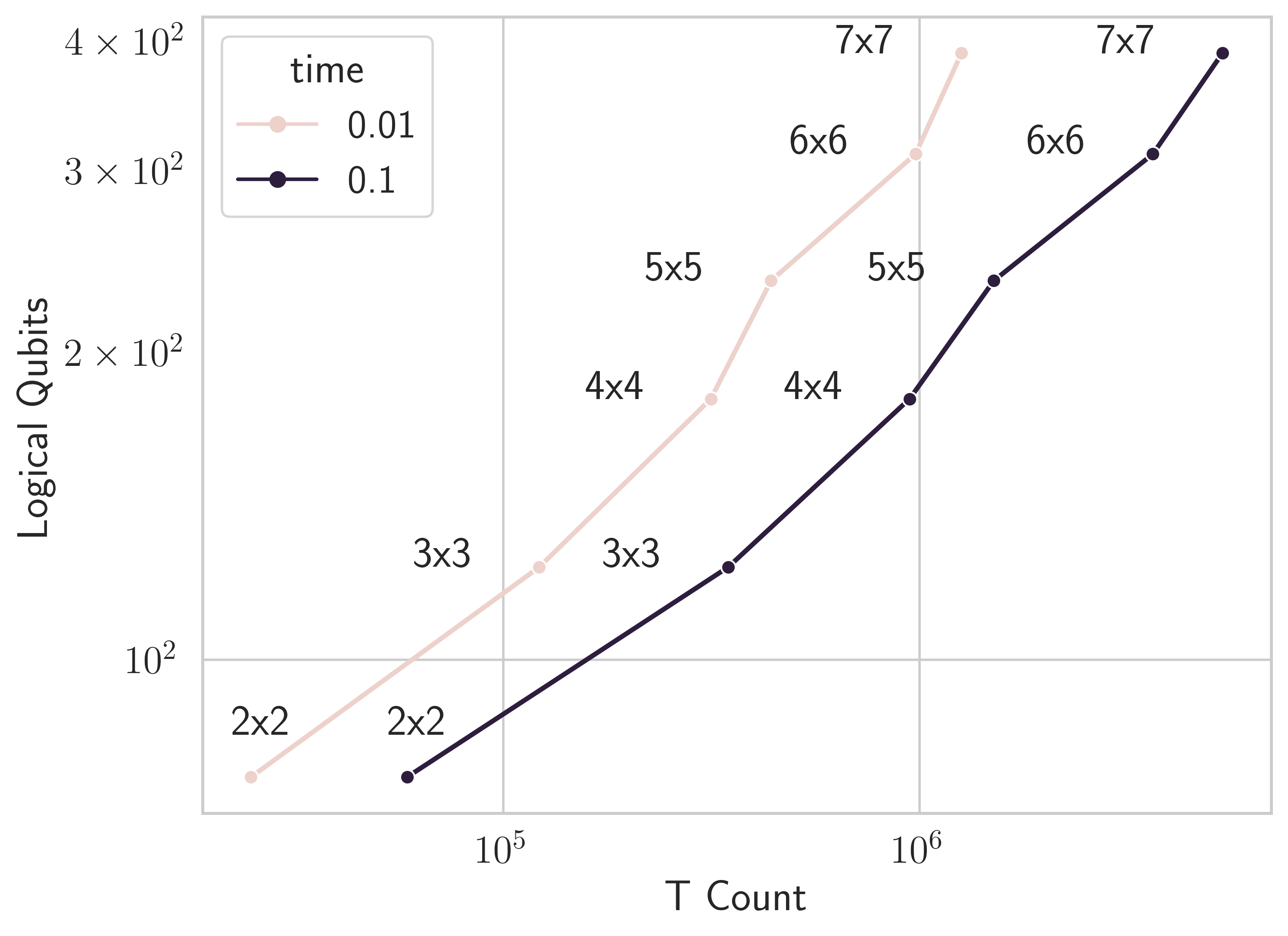}
    \caption{Logical resource scaling relationship for the time evolution portion of calculating dynamic correlation functions. The Hamiltonian considered is the Fermi-Hubbard model \cref{eq:fh_hamiltonian} with $U=2, V_{nn}=1, \mu=1$ and various lattice sizes.}
    \label{fig:logq_vs_tcount}
\end{figure}

In Fig.~\ref{fig:logq_vs_tcount} we plot the relationship between the number of logical qubits and the T count. 

The problem instances that resource estimates were generated for are limited to relatively small sizes, the largest being 7 by 7.
This is due to the fact that graph-state compilation typically carries out computationally expensive operations.
In order to compile larger circuits and to eventually run large-scale quantum computations, improvements to such compiler tools need to be made.
An alternative tool that has been developed for large-scale graph state compilation is the RubySlippers compiler~\cite{benchq}. Using such a tool would enable compilation of larger instances, though the resource estimates would likely be higher.
This points to the need for further development of logical architecture compilation tools.

% \documentclass{article}
% \usepackage{graphicx} % Required for inserting images
% \usepackage{physics}
% \usepackage{hyperref}
% \title{utility}
% \author{joshjob42 }
% \date{May 2024}

% \begin{document}

% \maketitle

\section{Utility estimation}
Utility estimation is a complicated subject. 
Some impacts, for instance, directly replacing existing HPC time with a better algorithm (either by solving larger problems that currently are intractable or solving existing problems better) is fairly straightforward to identify and estimate. 
Others, like technology spillover, are much more uncertain. 

For our purposes here, we choose to construct a probabilistic graphical model (PGM) for our utility estimates, to attempt to render as clear as possible the structure of our assumptions and uncertainty and give parameters for that model based on reference to the literature in physics and economics, budget information, and expert opinion (with supporting arguments if available). A PGM is a model describing the conditional probabilities of various variables in terms of other variables, which may themselves have distributions that are functions of yet other variables, and so on. Given a PGM, some data upon which to condition the model, and prior distributions for the various free parameters, one can then apply any standard software for sampling from them to produce estimates of the posterior distribution for the variables. By describing our utility estimates in terms of PGMs, then, we enable the reader to change some parameters or components should they disagree with our assessments.

We will categorize sources of utility in two ways: first order effects, and second order effects. 
The first order effects are the impacts that arise directly from the existence of an efficient Fermi-Hubbard solver. 
These incorporate the value of knowledge generation by such a tool e.g.~savings by using the tool to guide experimental search, and the reduced cost of generating said knowledge. 
The second order effects are the impacts that arise indirectly, e.g.~via technology spillover and human capital accumulation.

Since such a solver does not yet exist, except for the t-U model at half-filling \cite{PhysRevB.94.085103} where Auxiliary-Field Quantum Monte Carlo (AFQMC) is numerically exact and efficient, we begin our analysis based on the assumption that our solver's utility will derive from it being an improvement over existing solvers will thus redirect existing resources used on these sorts of simulations. 

One can consider a few basic utility thresholds.
\begin{enumerate}
    \item  Solving spin-1/2 $4\times 4$ Fermi-Hubbard lattices with nearest neighbor interactions for arbitrary parameters of $V_{nn},U$ to very small error ($\sim{10^{-12}}$). These lattices are exactly representable in 32 qubits, and can be solved exactly on large workstations. (In fact, truly arbitrary lattices can be considered at this scale.) Below this, no utility is possible.
    \item Solving this same class of problems but at $8\times 8$. Density Matrix Renormalization Group (DMRG) algorithms are routinely  applied on up to $8\times L$ lattices with closed/open boundary conditions such that the system forms an open cylinder, which are then used to project to the thermodynamic limit, as in \cite{PhysRevX.5.041041} and \cite{doi:10.1073/pnas.2109978118}. Runtimes for estimates of the ground state in this regime classically are on timescales of $\sim{10^5}$ core-hours, basing estimates on node sizes and runtimes discussed in \cite{PhysRevX.5.041041}.
    \item Solving this class at scales of $20\times 20$, arguably the smallest sizes to yield potential insights into the superconducting phase, based on an estimate of needing at least six full lattice units in real materials to observe the full correlation length of the system \cite{PhysRevB.70.104515}, and thus is likely the smallest scale to being to give real insights into the design of superconductors. Though it has recently been suggested \cite{PhysRevX.10.031016} that the ground state of this basic Fermi-Hubbard model is not superconducting, models at this scale cannot be studied except by techniques like AFQMC, which struggle away from half-filling due to the sign problem. Any reasonable results at this scale are likely to be interesting. Moreover, machines that can solve this model at this scale will also generally be able to solve other models such as the triangular lattice and t-J-J' model at scales larger than $10\times 10$, and even the 5-orbital model based on LaOFeAs \cref{eq:multiorb} 
    at $9\times9$, which is larger than any previous study we found. It would also be able to readily solve problems commonly modeled with approximate techniques at high computational cost today, such as the $8\times 48$ model studied at half-filling using AFQMC in \cite{PhysRevX.5.041041}.
    \item Solving the 5-orbital class based on LaOFeAs \cref{eq:multiorb}, at scale 20x20. This model much more closely follows a real superconducting material at a scale likely to begin to capture the decay of correlation in the ground state. 
    \item Solving the 5-orbital class based on LaOFeAs \cref{eq:multiorb} at $100\times 100$ model. This is almost certainly much larger than necessary to capture the full correlation length of the ground state for this model, and a machine capable of solving this model with reasonable accuracy (say, error $<0.01$\%) would also be able to study similar models, such as a 1-orbital next-to-nearest neighbor model, triangular lattices, etc., also at this scale, which may capture important features of real materials. At this scale, the machine would almost certainly be able to form the backbone of a materials search for superconducting materials, and provide deep insights into the physics of bulk 2D materials. 
\end{enumerate}
Utility for each threshold rises from merely non-zero at stage 1 (assuming an extremely short $O\sim(1s)$ time to solution), to encompassing the scale of many existing Fermi-Hubbard computational studies at stage 2, to encompassing essentially all existing studies, even highly approximate/low-accuracy ones, on the Fermi-Hubbard model at stage 3, and then pushing into uncharted waters at stage 4 and beyond up to and including potentially hastening the discovery of a room temperature superconductor. For now, we will focus on estimating the utility from stage 2 to 3, ie the utility of being able to replace all existing computational tools at relevant scales for this problem, and then discuss the potential but more speculative utility at stage 3 and above.

\subsection{A brief aside on the returns to science and discount rates}
It has been estimated by economists that funding for basic science research yields long-run present value returns to the economy of conservatively \$5 per marginal dollar invested \cite{jones2021science}. We won't address the methodology here, but \cite{jones2021science} gives a fairly thorough review of the literature and demonstrates nearly every estimate of investment in R\&D to yield much higher social rates of return than investments in virtually any other area of the economy. We here then take as our ground assumption that funding for science is more or less fixed, largely by political concerns, and so money saved in one area or field will end up reinvested in other fields. In that sense, a \$1 saved is equivalent to a \$1 invested in R\&D, and will return ultimately \$5 in net present value to the economy.

In addition, for savings/utility that accrue annually, for instance by reducing an annual expense, the net present value of all the future earnings must be estimated. This is done by calculating a present value assuming a discount rate, $\lambda$, which for our purposes we take as 5\%. Thus, earnings N years from now are worth $(1-\lambda)^N$ \$ in the present day. This reflects that people do not place as much value on a dollar earned in the future as they do today. The choice of discount factor effects to final outcome, but largely in a linear fashion, as the sum total of present value of all future earnings amounts to $1/\lambda$ times the annual return. A discount factor of 5\% is commonly used by governments in determining the value of a project, and is higher than the 10-year treasury bond yield for all but brief moments over the last twenty years. Thus, a 5\% is likely higher than optimal as a discount factor from the perspective of the government.

\hypertarget{first-order-effects}{%
\subsection{First order effects}\label{first-order-effects}}

\textbf{Value of reduced cost of data generation} We can assign a value by comparing with the cost of generating electronic structure for strongly correlated materials on classical computers. 
To do so, we will quantify the operating costs, power usage, and carbon footprint of computing facilities.

\begin{itemize}
% \tightlist
\item
  \textbf{Operating costs} -- The DOE estimates it spends \$586 million on operating costs of all high performance computing facilities. (This does not account for capital costs, however.)
  We estimate a significant fraction, approximately 20\%, corresponds to running this kind of material science problem, with a Beta distribution Beta(2,8) distribution (expectation is 20\% with a 67\% chance it is between \textasciitilde8\% and 31\%). This is consistent with findings from LANL that approximately 20\% of their HPC resources are used on modeling Fermi-Hubbard type systems. We estimated at approximately 10\% of the awards in the \hyperlink{https://doeleadershipcomputing.org/awardees/}{INCITE} program, were dedicated to solving electronic structure problems of the general kind described here (for various models), broadly in line with our distribution above.
\item
  \textbf{Energy usage} -- In FY21, LBNL used 41 million kWh of electricity to run NERSC. Assuming an approximate \$50/MWh price for electricity (approximately the current nationwide average price for bulk power from natural gas plants, the most common source of energy), this corresponds to $\$2.05M$
\item
  \textbf{Carbon footprint} -- In FY21, LBNL computational facilities emitted 13k MTCO2e (Metric Tons of CO2 equivalent). 
  Mutliplying by the estimated social cost of carbon, which in 2022 as approximately \$120/tonne (and up to 3x that roughly depending on discount rate choice), \cite{socialcostofcarbon}, this corresponds to $1.56M$ per year.
\end{itemize}

Combined, we then estimate that the annual utility just from being able to offload these costs from HPC environments can be computed easily, which kicks in at our stage 2 utility threshold of $8\times 8$ Fermi-Hubbard lattices and reaches 100\% by stage 3. Our total pool of costs from the operation of HPC facilities at DOE is \$589 million a year, rounding to the nearest million. We simply multiply that by the fraction of those resources being used for Fermi-Hubbard type applications to get a distribution for the potential annual value of offsetting computational costs at DOE facilities presently allocated to Fermi-Hubbard type simulations (the expectation value is approximately \$117M/yr).

\hypertarget{value-of-guiding-experimental-searches-by-inclusionexclusion-of-regions-in-parameter-space}{%
\subsubsection{Value of guiding experimental searches by inclusion/exclusion of regions in parameter space}\label{value-of-guiding-experimental-searches-by-inclusionexclusion-of-regions-in-parameter-space}}

We hope that simulations of Fermi-Hubbard problems can give insight into which regions of parameter space for new materials are potentially fruitful to search with costly experimentation. 
This will, however, generally require greater fidelity to real materials than the simple model $8\times 8$ or even standard $20\times 20$ lattice site utility thresholds (stage 2 and stage 3) permit. 
To truly deeply integrate with experiment, we expect that we will have to achieve our higher utility thresholds of stage 4 or 5.
We again consider three primary resource costs for experimental work: operating costs, energy usage, and carbon footprint.
We obtain numbers for some of the big ticket items where numbers are readily available: user facilities run by the DOE, specifically looking at Lawrence Berkeley National Laboratory (LBNL).

\begin{itemize}
% \tightlist
\item
  \textbf{Operating costs} -- The DOE estimates it spends \$525 million on operating costs of all their X-ray sources, and \$292 million on operating costs of all their neutron sources, \cite{DOE_budget_2023}.
\item
  \textbf{Energy usage} -- In FY21, LBNL used 55 million kWh of electricity to run the accelerator and laboratories on its campus.
\item
  \textbf{Carbon footprint} -- In FY21, LBNL experimental facilities emitted 19.2k MTCO2e (Metric Tons of CO2 equivalent).
\end{itemize}

If 1\% of these resources can be rededicated to another problem due to the existence of a Fermi-Hubbard solver, we would save an estimated \$8 million, 0.5 GWh, and 1.9 MTCO2e. 
We use a Beta(1,99) distribution to approximate the uncertainty here (expectation value 0.01, 95\% confidence interval (0.25\%,5.6\%)), for we have very limited visibility into the potential upside here and wish to be conservative. This is much more speculative and higher variance than the estimates based on supercomputer use above due to the lack of available information on what fraction, precisely, of these resource are centered on materials that can be modeled well with a Fermi-Hubbard solver. However, as an upper bound, we certainly cannot save more than the sum of x-ray and neutron source operating costs. In expectation, this contribution is taken as \$8M/yr.

\hypertarget{value-of-knowledge-generation}{%
\subsubsection{Value of knowledge
generation}\label{value-of-knowledge-generation}}

Over the past 2 decades, the Department Of Energy (DOE) has had 4000+ publications from the past 20 years in its OSTI database of funded published papers specifically focusing on the Fermi-Hubbard model.
Similarly, the National Science Foundation (NSF) has over 9000 publications focusing on strongly correlated materials. 
These indicate an immediate value -- both in the existence of a Fermi-Hubbard solver, and in the downstream impacts on a very interested community.

To put a value on this idea, we can refer to the budget of the DOE. 
The FY2023 budget report of the Department Of Energy indicates that \$202 million is spent on funding condensed matter and materials research, and \$13 million specifically in computational materials sciences \cite{DOE_budget_2023}. 
Thus, this gives us an idea for the scope of impact. 
Solving these types of Fermi-Hubbard instances should compose approximately 10\% of this (or \$1.3M).

There is also a question of potentially increasing the productivity of researchers in the field, by offsetting some of their time by employing a general purpose high-accuracy solver for the largest size models they consider currently at great expense and time costs.

To model this, we will estimate the size of the field itself by doing a literature search for the Hubbard model on OpenAlex \cite{priem2022openalex}, a free and open source literature database, yielding a total of approximately 98k papers. Isolating to the last ten years, from 2014 onward, we can then pull the number of works published by $10^4$ randomly selected authors of a paper in our dataset, and use as a proxy of their time the assumption each publication required approximately the same effort (a terrible approximation individually but over ten thousand authors and similar number of publications, fine enough for our purposes). For each author we compute what fraction of their research effort went into papers on Hubbard models, and then perform a Bayesian bootstrap over our ten thousand authors to compute an estimate for the average fraction of each author's time went into Hubbard model research. This leaves us with an estimate well approximated by Normal(0.105,0.0022), ie that approximately 10.5\% of their research output was focused on the Hubbard model. Using the average number of authors of papers published in the last decade as a guide and assuming an approximate average annual cost of salary and benefits of \$100k/yr, approximately the cost on a grant of a postdoc, we can then estimate a potential cost of \$$20.5\pm0.5$/yr for researchers time and effort. 

Together then, we can very roughly estimate approximately \$22M/yr for personnel collectively from the two above sources. Our uncertainties from the above will be swamped by uncertainty around how much productivity might be improved with access to efficient, fast, exact solvers. We will approximate the productivity gain to a distribution of Beta(1,4) (a mean of 20\% improvement). The resulting savings/improvement has a distribution with expectation value of \$4.4M and 10th, 50th, and 90th percentiles of \$0.6M,\$3.5M, and \$9.7M respectively.

\hypertarget{second-order-effects-1}{%
\subsection{Second order effects}\label{second-order-effects-1DOE_budget_2023}}

Here we will outline some additional sources of utility. However, since we are extremely uncertain about the concrete variables as these are longer term potential spinoffs, and also entangle with other problem instances, we will not include all of these potential sources of value and break out our resource estimate into two parts, inclusive and exclusive of the below.

\hypertarget{technology-spillover}{%
\paragraph{Technology spillover}\label{technology-spillover}}

These can arise from the development of new technologies that rely on the Fermi-Hubbard solver in order to generate the requisite knowledge. 
In principle, these are separate instances, but here we outline a few potential developments. Most, however, depend on or originate from the development of room-temperature reasonably high field and critical current superconductors. Estimating the probability of developing them depending on the existence of good Hubbard model solvers at our stage 4 and 5 utility thresholds is, fundamentally, a guessing game, contingent on expert opinion. They may be possible to develop without such solvers, but might be developed faster with one.

Here, we attempt to provide a model to describe this uncertainty, and leave it up to the reader to adapt the parameters to suit their own understanding and expectations. 

We have a few key parameters: 
\begin{enumerate}
    \item  The probability such a superconductor exists at all which we take as 80\% 
    \item What year such a superconductor would be found conditional on it existing, which we take as a log-normal distribution LogNormal(3.5,1), with [0.1,0.5,0.9] quantiles of [9.1,33.1,119.3] years respectively (parameters chosen to approximately match ~10 years, 30 years, and 100 years for these 3 quantiles)
    \item By how much would a solver accelerate development of a superconductor? For this, we assume it will shorten development time by some multiplicative factor, taking Beta(2,4) as our initial guess, which has an expectation value of 1/3, ie that with a stage 5 threshold solver, progress would occur approximately three times faster than otherwise.
    \item In what year will such a solver be developed? For planning purposes we will assume ten years from now.
\end{enumerate}

Combining these, we can get an estimate for the reduction in the time to development of a high-T, high-current, room temperature superconductor, the [0.1,0.5,0.9] quantiles of which are [0,12.3,60.3] years. Given a pair of years $y,z$ where $y$ is the year of development without our solver, and $z$ is the year of development with, using a discount rate $r$ and our 80\% probability of a solver existing at all, then the net present value of our solver in units of the annual savings today if we had such a superconductor is $\frac{0.8}{r} (1-r)^z - (1-r)^y$. Assuming a 5\% discount rate, this yields a net present value has a distribution with mean of 2.4, and 10th, 50th, and 90th percentiles of [0,2.2,5.2] years of present-day savings respectively. Thus, for each application below we can multiply by this distribution to get a distribution for the utility.

\begin{enumerate}
\def\labelenumi{\arabic{enumi}.}
% \tightlist
\item
  \textbf{More efficient/powerful MRI Machines:} A good MRI scan can save a life. 
  These machines are a very important part of medical procedures that save many lives. 
  They use high magnetic fields, which tend to require superconductors, and those operate at low temperatures which require liquid helium. 
  Due to helium shortage and size constraints, researchers are looking for helium free and low field imaging solutions. 
  In 2021, estimated domestic apparent consumption of Grade-A helium was 40 million cubic meters (1.4 billion cubic feet) \cite{staffusgshelium}, and it was used mostly (about 50\%) for magnetic resonance imaging in industry, medicine and laboratory. 
  An efficient Fermi-Hubbard solver can speed up the search for superconductors that do not require low temperatures and sustain higher currents. In turn, this can reduce helium usage or enable larger MRI machines so as to accommodate patients better and/or achieve higher resolution. The value of this contribution is largely unknown, though the MRI market is presently estimated to be approximately \$8B/yr \cite{MRImarket}. For now though, we will not include an explicit estimate for this in our final utility calculation.
\item
  \textbf{Power consumption:} As mentioned above, most of the energy used by a quantum computer is to keep the device cool. 
  High temperature superconductors can help us make quantum computers more practical for everyday use. This becomes somewhat circular however -- the value of this is tied to the value of quantum computers which we are in turn suggesting is tied to this, and so on. 

  More concretely, approximately 5\% of energy produced is wasted as heat in transmission and distribution \cite{EIAfaq} with perhaps 2-4\% of that coming from the transmission itself, amounting to approximately 80-160TWh of energy per year in the US, at a cost of approximately \$4-8B at present wholesale electricity prices of approximately \$50/MWh. Discounting the costs of installing superconducting cable (which would eat into the value of this proposition but could vary wildly), we can estimate this is part of the utility of our FH solver, and simply note that the cost to install cabling for transmission will in essence be deducted from this utility.
\item
  \textbf{Battery technology:} Improving the energy density of lithium-ion (Li-ion) batteries enables applications in electric vehicles and energy storage at an affordable cost. 
  Over the past ten years, however, innovation has stalled---battery energy density improved 50 percent between 2011 and 2016, but only 25 percent between 2016 and 2020, and is expected to improve by just 17 percent between 2020 and 2025, \cite{mckinsey_pervoskite}. 
  Recent research, \cite{kim2022fault}, has shown that quantum computing will be able to simulate the chemistry of batteries in ways that can't be achieved now. 
  Quantum computing could allow breakthroughs by providing a better understanding of electrolyte complex formation, by helping to find a replacement material for cathode/anode with the same properties and/or by eliminating the battery separator. 
  As a result, we could create batteries with 50 percent higher energy density for use in heavy-goods electric vehicles, which could substantially bring forward their economic use, \cite{mckinsey_pervoskite}. 
  However, again, how large a role solving Fermi-Hubbard would be in this is unknown and unestimatable in our opinion at this time.
\item
  \textbf{Material Search and related technologies:} Fermi-Hubbard type solver can enable us to simulate the properties of materials and predict application based usage. 
  There are many strongly correlated materials under research where scientists are trying to get the predictions for properties at different parametric and environmental conditions. 
  For instance, quantum spin liquids are believed to make an efficient information storage device, show interesting magnetic properties, make practical candidates for topological qubits and others, \cite{savary2016quantum}.
  There are some organic materials which can be used in biomedical research and are among strongly correlated materials. 
  Considering the active research in materials science we can find many instances where having a Fermi-Hubbard solver can be useful, \cite{coulon1987magnetic}.
\end{enumerate}

While we outlined a number of potential sources for utility above, we will here only take a fairly conservative one -- that should we develop room temperature high-field, high critical current superconductors that we will ultimately deploy them in transmission lines across the US. This is, as we said, quite conservative -- such a material would also be useful in fusion, maglev trains, consumer electronics, medical imaging, sensing, among other areas. We thus believe that this readily estimatable utility of such a superconductor should serve as a lower bound on the likely ultimate utility of its development.

Using a uniform distribution on the \$4-8B/yr current annual cost of electricity lost in transmission lines, we can then convolve that distribution with our above estimate of the net present value of our stage 4/5 utility threshold Hubbard solver to get a a distribution for the net present value of such a solver in solving the superconductor problem. This distribution has a mean of approximately \$10B, with 10th, 50th, and 90th percentiles of \$0, \$9.2B, and \$21.6B (given our assumptions, there is a greater than 10\% chance in our model that such a superconductor will be developed before our Hubbard solver).

\hypertarget{human-capital-accumulation}{%
\subsubsection{Human capital accumulation}\label{human-capital-accumulation}}

The existence of a new technology drives a demand for knowledgeable users of said technology. 
From the academic perspective, this can come in several forms:

\begin{itemize}
% \tightlist
\item
  Higher demand for end-users of the Fermi Hubbard solver in industrial settings
\item
  Higher demand for end-users leads to a need for educational programs, growing the need for faculty, expert users, and degree/certificate programs.
\item
  Access to new tools enables career advantages for users of the technology, both tangible and not. 
  Translating this to concrete numbers is difficult, as we would need estimates of the size of these populations across academia and industry, which is quite difficult, and estimate the opportunity costs associated with workers going into quantum science specifically versus their next best alternatives (which itself produces a kind of self-referential quality, as that in part is contingent on the value of quantum computing to the economy, which is what we are trying to estimate). 
  Thus, we will not quantify this specifically at this time.
\end{itemize}

\hypertarget{final-utility-estimate}{%
\subsubsection{Final utility estimate}\label{final-utility-estimate}}

\begin{figure}
    \centering
    \includegraphics[width=\columnwidth]{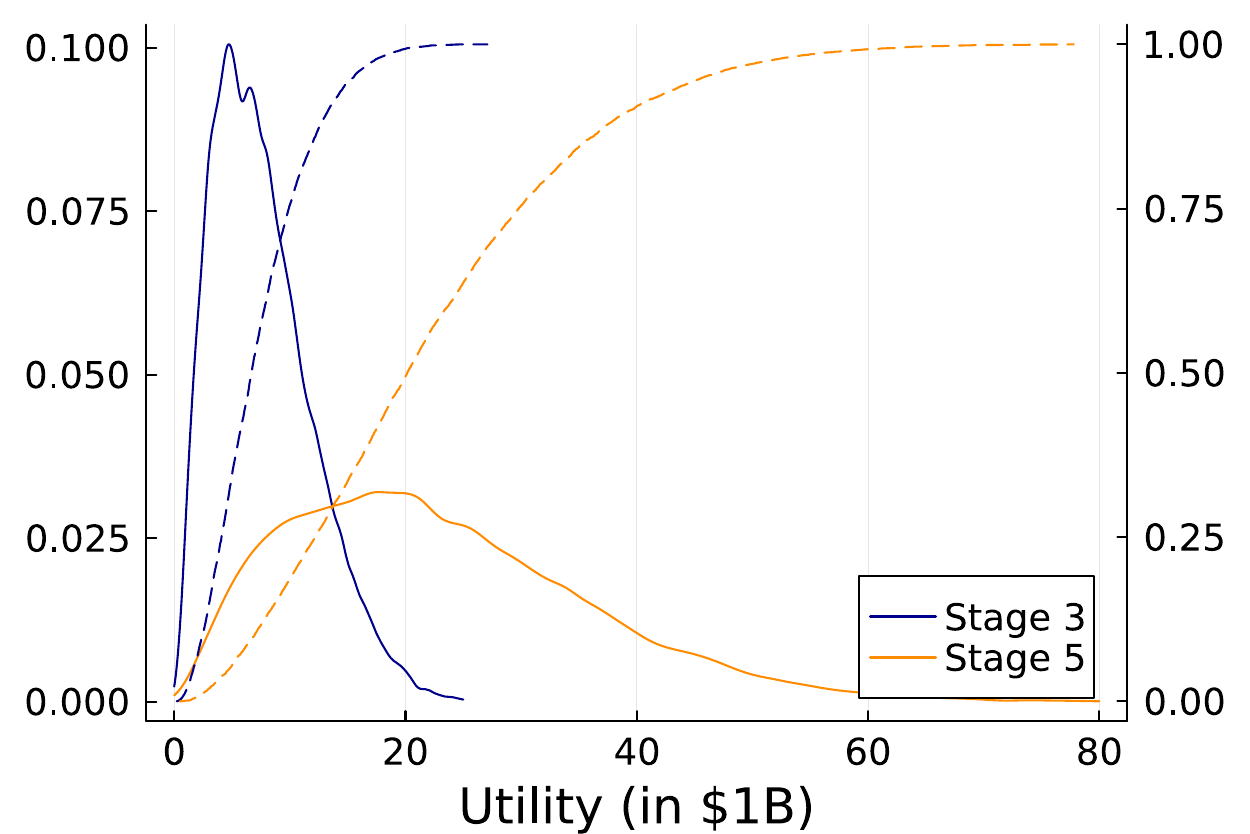}
    \caption{Probability density function (PDF, solid) and cumulative distribution function (CDF,dashed) for our estimates for the utility of a machine that can solve the stage 2/3 models (blue) and stage 4/5 models (orange), in units of \$1B.}
    \label{fig:utility_estimate_CDF}
\end{figure}

Combining the first order effects with associated uncertainties discussed above, we arrive at a mean estimate for the total net present value of a machine able to solve these problems constructed in 2034 given a 5\% discount rate, constant annual return at 2022 rates, and the net future effects on economic growth, assuming a conservative 5x return due to redirecting spending to other scientific fields and that spending's effect on the economy long
term. For stage 2-3 utility thresholds, ie replacing all existing workloads for Fermi-Hubbard models, our expectation value is \$7.3B. Annual direct return ignoring redirection of funds works out to 1\% of that number, \$73M/yr in present value (ie approximately \$128M/yr starting in 2034). Stage 2 will likely not capture all of this value, but stage 3 certainly does.

For stage 4-5 utility thresholds, ie sufficiently large to guide experiments and potentially hasten the development of a room temperature superconductor, we estimate a mean, sans superconductor, of \$7.8B. Including potential returns from more efficient transmission alone from a room temperature, high-field, high critical current superconductor, we estimate a mean of \$22.1B of present value.

Our probability and cumulative density functions for stage 2/3 and 4/5 are plotted in \cref{fig:utility_estimate_CDF}.

% \end{document}

\section{Conclusion}

In this work, we have outlined the case for the \FH model
as a scalable benchmark for quantum
computers.
In short, it is a model where the base
case is well known, and that exhibits sufficient
complex physics to be of interest to the
wider community. It is extensible in a number
of directions (e.g. lattice and orbital complexity), and it yields enough complex
physics that it can be ``difficult'' from
the perspective of classical computing.

Moreover, it subsumes enough
of the current efforts to be of value from the
initial use of a putative \FH model solver.
To quantify the latter,
we have
outlined the utility of
having a device (quantum computer or otherwise)
that can provide experimentally-relevant
quantities based on its first order effects
(financial impact, energy impact, and carbon
footprint impact),
and potential second order effects
including impact on society through
technological advancements if such a solver
could lead to room temperature superconductivity.

In order to demonstrate how it can be used
as a benchmark, we have performed the first of 
its kind
detailed resource breakdown for the task of estimating experimentally-relevant dynamical correlation functions of the Fermi-Hubbard model.
We have estimated the number
of logical qubits, T-gate counts, bus size and runtime for
the time evolution portion of this problem,
running on a two-fridge measurement-based quantum computer.
Rather than a high level view,
we performed a detailed estimate that
takes the full process into account,
including the full cost of T state distillation,
(bus) ancilla qubits.
These estimates were based on a particular
choice of time evolution 
algorithm (QSP), error correcting
code (surface code), and we expect the precise
results to vary as new algorithms and 
error correcting codes are developed.
However, based on the current methodology,
we find that running the time evolution up to a modest time for a $7\times 7$ cluster
requires a calculation that takes over 3 years,
and uses 400 logical qubits. Given that a $7\times 7$ cluster is at the extreme end of
today's classical hardware, this points to a
need for improvement along a variety of axes.

The clear need for improvement that comes out
of the resource estimation is precisely why
scalable, extensible benchmarks are needed.
In order to make the case that a particular
computational tool is going to make a difference,
we should have a clear view of the costs
associated, and the potential benefits it
may have. The \FH model provides a clean
platform to lay out this case.

\section*{Acknowledgements}
This work was supported by the
Defense Advanced Research Projects Agency (DARPA)
under Contract No. HR001122C0063.  J.E. and K.M. also specifically acknowledge that this material is based upon work supported by the Defense Advanced Research Projects Agency under Air Force Contract No. FA8702-15-D-0001. Any opinions, findings, conclusions or recommendations expressed in this material are those of the author(s) and do not necessarily reflect the views of the Defense Advanced Research Projects Agency. 

\bibliography{refs}

\end{document}